\documentclass{article}
\usepackage{arxiv}
\usepackage{amsthm}
\usepackage{amsmath}
\usepackage{natbib}
\usepackage[colorlinks,citecolor=blue,urlcolor=blue,filecolor=blue,backref=page]{hyperref}
\usepackage{graphicx}

\numberwithin{equation}{section}
\theoremstyle{plain}

\theoremstyle{plain}

\usepackage{amsmath, bm, amssymb, dsfont}
\usepackage{booktabs} 
\usepackage{subcaption}
\usepackage{graphicx}
\usepackage{pgfplots}
\usepackage[all]{nowidow}
\usepackage[utf8]{inputenc}
\usepackage{tikz}
\usetikzlibrary{er,positioning,bayesnet}
\usepackage{multicol}
\usepackage{multirow}
\usepackage[linesnumbered,ruled,vlined]{algorithm2e}
\SetKwInput{KwInput}{Input}                
\SetKwInput{KwOutput}{Output}              
\usepackage{mathtools}
\usepackage[inline]{enumitem} 

\usepackage{xcolor}
\usepackage{comment}
\pgfplotsset{compat=1.18}
\usepackage{float}
\newcommand{\iid}{\stackrel{\mathrm{iid}}{\sim}}
\newcommand{\wrt}[1]{\mathrm{d}{#1}}
\newcommand{\ind}{\stackrel{\mathrm{ind}}{\sim}}

\usepackage{amsmath}

\usepackage[normalem]{ulem}

\usepackage{natbib}
\setcitestyle{authoryear,open={(},close={)}}

\begin{document}

\title{Personalized Treatment Selection via Product Partition Models with Covariates}
\date{}
\author{ 
Matteo~Pedone \\
	Department of Statistics, Computer Science and Applications\\
	University of Florence \\
	\texttt{matteo.pedone@unifi.it} \\
	\And
	Raffaele Argiento \\
	Department of Economics\\
	University of Bergamo\\
	\texttt{raffaele.argiento@unibg.it} \\
	 \AND
Francesco C.~Stingo \\
	Department of Statistics, Computer Science and Applications\\
	University of Florence \\
	\texttt{francescoclaudio.stingo@unifi.com} \\
}
\maketitle

\begin{abstract}
Precision medicine is an approach for disease treatment that defines treatment strategies based on the individual characteristics of the patients. Motivated by an open problem in cancer genomics, we develop a novel model that flexibly clusters patients with similar predictive characteristics and similar treatment responses; this approach identifies, via predictive inference, which one among a set of treatments is better suited for a new patient.
The proposed method is fully model-based, avoiding uncertainty underestimation attained when treatment assignment is performed by adopting heuristic clustering procedures, and belongs to the class of product partition models with covariates, here extended to include the cohesion induced by the Normalized Generalized Gamma process.
The method performs particularly well in scenarios characterized by considerable heterogeneity of the predictive covariates in simulation studies. 
A cancer genomics case study illustrates the potential benefits in terms of treatment response yielded by the proposed approach. Finally, being model-based, the approach allows estimating clusters' specific response probabilities and then identifying patients more likely to benefit from personalized treatment.
\end{abstract}

\maketitle

\section{Introduction}
\label{sec:intro}
Cancer comprises a collection of complex diseases characterized by heterogeneous cellular alterations across patients and cancer cells within the same neoplasm \citep{bedard2013tumour}. Patients with similar clinical diagnoses may show diverse responses to the same treatment due to tumor heterogeneity. A treatment for a particular diagnosis may be effective on average, but its effectiveness may vary across subpopulations. 
In recent years many attempts have been made to devise personalized treatment strategies that leverage patients' characteristics, including the tumor's genome, to identify the treatment with the highest likelihood of success \citep{simon2010clinical}.  
Within this precision medicine paradigm, there is an increasing interest in discovering individualized treatment rules (ITRs) for patients that show heterogeneous responses to treatment, e.g., when the treatment effect varies across groups of patients. An ITR is a decision rule that assigns the patient to the treatment given patient/disease characteristics \citep{ma2015statistical}. The optimal ITR is the one that maximizes the population mean outcome.
Statistical methodology research in precision medicine is devoted to developing personalized treatment rules to inform decision-making. The distinctive mark of statistical inference under the precision medicine paradigm is to leverage heterogeneity to improve therapeutic strategies \citep{kosorok2019precision}. 

Our interest specifically lies in developing frontline treatment selection rules rather than estimating treatment's causal effects, as commonly done within the ITR framework.
Conventional methods for treatment selection rules are based on semi- and non-parametric procedures to identify subgroups of patients more likely to benefit from a treatment leveraging few baseline markers \citep{bonetti2000graphical, song2004evaluating}. The subgroup approach can provide valuable information when performed according to a prespecified analysis plan. Nonetheless, stratified subsets of patients defined by one or few biomarkers are often inadequate to account for patient heterogeneity and ultimately fail to establish effective treatment selection rules \citep{pocock2002subgroup}.
Other approaches account for patient heterogeneity by including covariates \citep{zhang2012robust, zhao2012estimating}. However, for these methods, the correct definition of treatment-by-markers interactions is crucial and relies on sensitive assumptions, which are difficult to specify in the clinical practice and may be limited to generalized linear models \citep{ma2016bayesian}.

To overcome these limitations, \cite{ma2016bayesian, ma2018integrating, ma2019bayesian} have established a hybrid two-step predictive model for personalized treatment selection. In the first step, a clustering algorithm based on a pre-defined genomic signature (predictive markers) is used to obtain a heuristic measure of the patients' molecular similarity. In the second step, given this measure of patients' similarity and a set of prognostic markers, a Bayesian model is used for treatment selection; specifically, for a new, untreated patient, the model predicts the treatment response probabilities for each competing treatment. This framework establishes two significant improvements over existing methods. Firstly, the common assumption of statistical exchangeability among patients is relaxed. Since each tumor is unique, patients are considered partially exchangeable only to the extent to which their tumors are molecularly similar. Moreover, this approach utilizes complementary sources of information for treatment selection, integrating predictive and prognostic characteristics of a patient.

This paper proposes a Bayesian predictive model for personalized treatment selection that builds upon \cite{ma2019bayesian} and overcomes some of its main limitations. As in \cite{ma2019bayesian}, we leverage prognostic determinants and predictive biomarkers for treatment selection. 
We propose a fully Bayesian integrative framework for clustering and prediction that performs all inferential tasks in a single model avoiding multi-step procedures; the proposed approach results in a treatment selection rule that fully accounts for patients' heterogeneity. Note that in \cite{ma2019bayesian}, the patients' similarities were estimated in the first step and included as known quantities in the second step; moreover, in the first step, two arbitrary choices had to be made, namely the clustering algorithm and the number of clusters. The proposed method accounts for the uncertainty in all modeling steps, resulting in improved prediction performances. In particular, we use a product partition model with covariates \citep[PPMx,][]{muller2011product} to cluster observations that are similar in terms of the values of the predictive covariates; specifically, the predictive covariates enter the model through the prior for the random partition. The resulting partitions are only partially exchangeable, and patients with similar covariates are \emph{a priori} more likely to be clustered together. In this paper, we use the \emph{cohesion function} induced by the Normalized Generalized Gamma process (NGGP) as a building block of our PPMx model to mitigate the \emph{rich-get-richer} property of the Bayesian nonparametric (BNP) priors.
Namely, the \emph{rich-get-richer} is the tendency for a small number of clusters to become overrepresented as more data points are added to the process, resulting in few large clusters and potentially many singletons. Despite being well studied in the Bayesian nonparametric literature as a prior inducing a Gibbs-type random partition \citep{lijoi2007controlling}, NGGP still has no common use. 
To the best of our knowledge, this is one of the first attempts the NGGP is employed as \emph{cohesion function} in a PPMx model \citep[see][]{argiento2022clustering}. 

We devise a method that, given the patients' prognostic and predictive markers, assigns them to the treatment with the highest likelihood of positive response. 
Prognostic covariates influence disease progression regardless of the treatments given to the patient, whereas predictive covariates change the likelihood of a positive response to a particular treatment. 
Conceptually, our strategy for selecting the optimal treatment for the new, untreated patient can be broken down into three steps. First, we consider historical patients and cluster them separately for each treatment according to their predictive markers. In this way, patients that underwent the same treatment are divided into homogeneous clusters with respect to predictive biomarkers. Then, we compute the utility provided by each competing treatment to the new untreated patient by assigning the new patient to the subgroup of historical patients with whom he shows the largest similarity in terms of predictive markers. 
The utility function relies on the model's posterior predictive distribution, which depends on both prognostic and predictive biomarkers. Finally, we select the treatment that ensures the largest predicted benefit. 

We apply the proposed method to a brain cancer dataset \citep{ma2019bayesian}, comprising 158 patients equally assigned to either standard or targeted treatment. For each patient, prognostic and predictive biomarkers, both consisting of pre-selected genomics markers, are available in addition to their categorical response to treatment. To facilitate optimal treatment selection, we assign numerical utilities to each treatment response level. This leads to a median utility score, which serves as a one-dimensional criterion for treatment selection. Our model shows good predictive performances and provides a sound framework for the identification and interpretation of clusters of patients.
\section{Bayesian Integrative Model}
\label{sec:bim}
We consider $n$ historical patients treated with $T$ alternative treatments, whose predictive and prognostic biomarkers are measured along with a discrete set of response levels of the clinical outcome. 
Let $a = 1,\dots,T$ index treatments and $n = \sum_{a=1}^{T}n^a$ be the total number of treated patients, of which $n^a$ assigned to therapy $a$. 
Note that, in our notation the superscript $a$ is solely a treatment index.
The treatment response $y_{i}^{a}$ of patient $i$ is a categorical variable with $K$ levels that encodes the residual disease extent after a clinically relevant post-therapy follow-up period. In particular, $y_{i}^{a}$ follows a multinomial distribution $y_{i}^{a}|\bm\pi_{i}^{a} \ind\text{Multinomial}(1, \bm\pi_{i}^{a})$, for $i=1, \dots, n^a$, with associated probability vector $\bm{\pi}_{i}^{a} =(\pi_{i1}^a,\dots,\pi_{iK}^a)^\top$; $\pi_{ik}^a$ is the probability of observing outcome $k$ for the $i-$th patient under treatment $a$, for $k=1, \dots K$. 
These probabilities will depend on $\bm z^{a}_{i}$ and $\bm x^{a}_{i}$, the $P-$ and $Q-$dimensional vector of prognostic and predictive features measured on the $i-$th patient that received treatment $a$, respectively.
We assume that patients with similar predictive biomarkers and the same prognostic covariates will respond similarly to a given treatment. To quantify the effectiveness of each competing treatment for patients with similar values of the predictive biomarkers, we adopt a covariate-dependent random partition model (RPM). 
For each treatment $a = 1, \dots, T$, patients receiving treatment $a$ are partitioned into clusters based on their predictive biomarkers $\bm x^a$.
Namely, we make the random partitions depend on predictive biomarkers.
Section \ref{sec:bnp} will describe the covariate-dependent RPM we use to achieve this goal. In this section, we assume that $\mathcal{P}^{a}_{n^a}=\{S_1^a,\dots,S_{C^{a}_{n^a}}^a\}$ is a given treatment-specific partition of the indices $\{1,\dots,n^a\}$, where $C^{a}_{n^a}$ is the number of clusters among patients treated with therapy $a$ and $n^a_j = |S^a_j|$ is the cardinality of cluster $j$, for $j=1, \dots, C^{a}_{n^a}$. Since we will later treat the partition of the units as a random quantity, the partition itself and the number of clusters depend on the number of observations, $n^a$. 
Following a common convention, we identify cluster-specific quantities using the superscript ``$\star$''. For example, when considering cluster $S^{a}_{j}$, the response vector is $\bm{y}^{a\star}_{j}=\{y_{i}^a: i\in S_j^a\}$, while $\bm{x}_j^{a\star}=\{\bm{x}_i^a: i\in S^a_j\} $ is the partitioned covariate matrix. 
We define the following hierarchical model for the response variables:
\begin{equation}
    \begin{split}\label{eq:model0}
y_{i}^{a}|\bm{\pi}_{i}^a &\ind \text{Multinomial}(1,\bm{\pi}_{i}^{a}) \\
\bm \pi_{1}^a,\dots,\bm \pi_{n^a}^a &\mid\bm\eta_1^{a\star},\dots,\bm{\eta}_{C^{a}_{n^a}}^{a\star},\mathcal{P}_{n^a}^{a},\bm \beta \sim  \prod_{j=1}^{C^{a}_{n^a}}\prod_{i\in S^{a}_{j}}\text{Dirichlet}(\bm \gamma_i^a(\bm \eta_{j}^{a\star},\bm \beta)),
\end{split}
\end{equation}
where $\bm \gamma_{i}^{a}(\bm \eta_{j}^{a\star},\bm \beta)= (\gamma_{i1}^a( \eta^{a\star}_{j1},\bm \beta_1),\dots,\gamma_{iK}^a ( \eta^{a\star}_{jK},\bm \beta_K))^\top$ is a vector of log-linear functions of the prognostic markers and cluster-specific parameters defined as follows:
\begin{equation}\label{eq:loglinpred}
\log(\gamma_{ik}^a(\eta^{a\star}_{jk},\bm \beta_k))= \eta^{a\star}_{jk}+\beta_{1k}z_{i1}^a +\dots+\beta_{Pk}z_{iP}^a.
\end{equation}
Model (\ref{eq:model0}) 
is robust with respect to overdispersion \citep{corsini2022dealing}, which is usually observed in multivariate categorical data \citep{chen2013variable}.
Predictive biomarkers, $\bm x^a$, enter equation \eqref{eq:loglinpred} through the parameter vectors $\bm \eta^{a}_{1}, \dots, \bm \eta^{a}_{C^{a}_{n^a}}$ and the partition $\mathcal{P}^{a}_{n^a}$ that depends on the predictive covariates $\bm x^a$, as we will elaborate in Section \ref{sec:pd}.
The $K$-dimensional vectors $\bm \eta^{a\star}_1,\dots,\bm \eta^{a\star}_{C^{a}_{n^a}}$ are cluster-specific parameters; high values of $\eta^{a\star}_{jk}$ correspond to an high probability of observing response $k$ for an individual treated with treatment $a$ in cluster $j$. We enforce $\bm\eta_{j}^{a\star}$ to be treatment-specific, and, as a consequence, the partitions $\{\mathcal{P}^{a}_{n^a}\}_{a=1, \dots, T}$ are independent across treatments. This construction provides a comparison among competing treatments. In fact, it allows patients with close genetic profiles that received different treatments to have distinct response probabilities.
Finally, $\bm \beta=(\bm \beta_1,\dots,\bm \beta_K)$ is a $P\times K$ matrix of regression parameters shared across treatments. Prognostic biomarkers enter equation \eqref{eq:loglinpred} as linear terms. Since prognostic determinants impact the likelihood of achieving a given therapeutic response regardless of the treatment, the associated coefficients are defined across therapies. Thus, prognostic covariates set a baseline response probability measure. 
Since patients should not be regarded as statistically exchangeable with respect to predictive biomarkers \citep{ma2016bayesian}, we leverage predictive biomarkers to drive the clustering process within each treatment. The resulting cluster-specific parameters $\bm \eta^{a\star}_j$ assess the benefit offered by a specific treatment on groups of similar patients. 
Note that the linear predictor is a function of the prognostic biomarkers only: the predictive covariates enter non-linearly equation \eqref{eq:loglinpred} only through the cluster- and treatment-specific parameters $\bm \eta^{a\star}_j$. This construction results in a random intercept that estimates the adjustment provided by predictive biomarkers to the baseline prognostic response probability on account of groups of patients with close predictive determinants. 
Note that, while the Multinomial logit model \citep{agresti2019introduction} could have provided similar predictive performance, its interpretation would have been less straightforward since the parameters represent log odds ratios with respect to a specific baseline response level.
\section{Prior distributions}
\label{sec:pd}
We assume independent shrinkage priors for the parameters $\bm{\beta}_k$. 
In particular, we adopt horseshoe priors \citep{carvalho2010horseshoe}, which belong to the class of global-local scale mixtures of normals. More in details, for $p=1, \dots, P$ and $k=1, \dots, K$
\begin{equation*}
\label{eq:hs}
    \beta_{pk} \iid N(0,\lambda_{pk}^2\tau_{k}^{2}), \,\,\,\,
    \lambda_{pk}, \tau_k \iid HC(0, 1), 
\end{equation*}
\noindent
where $HC$ denotes a half-Cauchy distribution, $\{\lambda_{pk}\}$ are local shrinkage parameters, and $\{\tau_{k}\}$ are global shrinkage parameters. All coefficients will be nonzero, but only those supported by the data will have large values due to the heavy tails of the prior. The joint distribution of the clustering and the cluster-specific parameters $(\mathcal{P}_{n^a}^{a}$, $\bm \eta_{j}^{a\star})$, is assumed to be independent across treatments. Therefore we will omit the superscript $a$ throughout Sections \ref{sec:pd} and \ref{sec:bnp}. In particular, we assume a product partition model with covariates \citep[PPMx,][]{muller2011product}, that induces independence across clusters and conditional independence within clusters. We detail our proposal for the PPMx on $\mathcal{P}_{n}$ in Section \ref{sec:bnp}.
Here, given $\mathcal{P}_{n}$, we details the  prior for $\bm \eta_{j}^{\star}$, $j=1,\dots,C_n$. 
We assume conditional independence between clusters, that is $\bm\eta^{\star}_j\sim G_0$, for $j=1, \dots, C_n$, where  $G_0$ is a prior for cluster-specific parameters. Then, the joint law of $(\mathcal{P}_{n}, \bm \eta_{j}^{\star})$ is assigned hierarchically as: 
    \begin{equation*}
    \label{eq:hierppmx}
            \bm \eta_{j}^{\star}  \iid G_0,~\text{for}~ j = 1,     \dots, C_n,\,\,\,\,\,\,
            \mathcal{P}_n \mid \bm x  \sim PPMx(\bm x).
    \end{equation*}
Specifically, we take $G_0$ to be a $K-$dimensional multivariate normal distribution and assume that $\bm\eta^{\star}_j|\bm\theta, \bm\Lambda\iid N_K(\bm\theta_, \bm\Lambda^{-1})$. To achieve more flexibility, we add an extra layer of hierarchy by assuming $\bm\theta|\bm\mu_0, \bm\Lambda, \nu_0 \sim N_K(\bm\mu_0, (\nu_0\bm\Lambda)^{-1})$ and $\bm\Lambda|s_0, \bm \Lambda_0 \iid W(\bm \Lambda_{0}, s_0)$, where $W$ is a Wishart distribution, with mean $s_0\bm\Lambda_0$. As customary hyperparameter choice, we set $\bm \mu_0$ to be the $K-$dimensional vector of $0$, $s_0=K+2$, $\bm\Lambda_0$ to be a $K\times K$ diagonal matrix with elements on the diagonal being equal to $10$, and $\nu_0 = 10$. Elicitation for the latter two parameters is discussed in Supplementary Material A. 
\section{Bayesian Nonparametric Covariate Driven Clustering}\label{sec:bnp}
In this section, we introduce the Product Partition Model (PPM) and describe its extension to incorporate the Normalized Generalized Gamma process (NGGP). We follow \cite{muller2011product}'s approach to integrate predictive biomarkers into the model, making the random partition dependent on predictive markers. 
We devise a covariate-dependent prior on the random partition that enables predictive markers to drive the clustering process. Thereby, we induce clusters of homogeneous observations in terms of predictive biomarkers. The resulting model defines independence across clusters and exchangeability only within clusters. 
The joint evaluation of prognostic and predictive covariates guides the optimal treatment selection, our main inferential goal. Still, only the predictive markers identify patients likely to benefit from a particular therapy. 
In this way, we may quantify the extent of benefit offered by a specific treatment on groups of patients characterized by similar values of the predictive markers. 
We denote with $\mathcal{P}_n:=\{S_1, \dots, S_{C_n}\}$ the partition of the data label set $\{1, \dots, n\}$ into $C_n$ subsets $S_j$, for $j=1, \dots, C_n$ and with $n_j=|S_j|$ being the cardinality of cluster $j$. 
In the seminal paper by \cite{hartigan1990partition} the prior on $\mathcal{P}_{n}$ is assigned by letting 
\begin{equation}
p(\mathcal{P}_n)=V_{n,C_n}\prod_{j=1}^{C_n}\rho(S_j),
\label{eq:prodpart}
\end{equation}
where $\rho(\cdot)$ is referred to as cohesion function, and quantifies the unnormalized probability of each cluster \citep{muller2011product}. Moreover, $V_{n,C_n}$ is a normalizing constant assuring that the prior sum up to one over the space of all partitions of the integers $\{1,\dots,n\}$. 
If $\rho(S_j)$ is only a function of $n_j=|S_j|$, then the resulting model for $\mathcal{P}_{n}$ is invariant under permutations of the labels of the set of integers $\{1, \dots, n\}$. Under this assumption, the resulting model for $\mathcal{P}_{n}$ falls in the class of Gibbs-type priors \citep{gnedin2006exchangeable}. In this framework, the cohesion assume the analytical expression $\rho(S_j)=(1-\sigma)_{n_j-1}$ with $\sigma < 1$ and $(1-\sigma)_{n_j-1}$ being the rising factorials, defined as $(a)_n=a(a+1)\dots (a+n-1)$, with $(a)_0=1$;  $p(\mathcal{P}_n)$ is denoted as {\it exchangeable partition probability function} (eppf) and the normalizing constant $V_{n,C_n}$ must satisfy the triangular recursion $V_{n,C_n}=V_{n+1,C_n}(n-\sigma C_n)+V_{n+1,C_n+1}$ for each $n>1$ and $1\le k \le n$ with the proviso that $V_{1,1}=1$. 
Note that, since $\rho(S_j)$ is an increasing function of the cluster size $n_j$, heavily populated clusters are more likely. This leads to the \emph{rich-get-richer} behaviour in the clustering induced by the BNP prior.
The connection between product partition models and Gibbs-type prior has been deeply investigated since the seminal paper by \cite{quintana2003bayesian}, see also \cite{de2013gibbs}. In this paper we choose $\sigma\ge 0$ and introduce a new parameter $\kappa>0$ such that:
	$V_{n, C_n} = \dfrac{1}{\Gamma(n)}\int_{0}^{\infty}u^{n-1}exp\big\{-(1/\sigma)[(\kappa+u)^{\sigma}-\kappa^{\sigma}]\big\}(\kappa+u)^{-n+\sigma C_n}\wrt u$.
In this way, the law of $\mathcal{P}_{n}$ coincides with the one induced by the 
Normalized generalized gamma process \citep[NGGP, ][]{lijoi2007controlling}. The NGGP encompasses the well known Dirichlet process (DP) when $\sigma=0$. 
In particular, \cite{lijoi2007controlling} highlighted the role of $\sigma$ in the \emph{predictive mechanism} of an NGGP: 
$p(\tilde{\bm\eta}\in\cdot\mid{\mathcal{P}_n},\bm\eta_1^\star, \dots, \bm\eta_{C_n}^\star)=\dfrac{V_{n+1, C_n+1}}{V_{n, C_n}}
G_0(\cdot)+\dfrac{V_{n+1, C_n}}{V_{n, C_n}}\sum_{j=1}^{C_n}(n_j-\sigma)\delta_{\bm\eta_{j}^{\star}}.$ 
The above formula describes the rule used to assign a new observation to a cluster, where the summand on the left represents the probability of forming a new group, and the one on the right represents the probability of being assigned to an already observed group. 
It is apparent that larger values of $\sigma$ increase the probability of generating new groups. From our simulation study \citep[Supplementary Material B.1; see also ][ Section 3.2]{lijoi2007controlling} large values of $\sigma$ also reduce the number of estimated singletons. These behaviours result in mitigating the rich-get-richer property of BNP priors.
We also assume a discrete prior distribution for $(\kappa, \sigma)$. In this way, we let the data choose the appropriate reinforcement rate \citep{lijoi2007controlling}, and we overcome a critical  ``trade-off" occurring when $\kappa$ and $\sigma$ are set to a fixed value. Indeed, both the parameters $\sigma$ and $\kappa$ have an effect on the number of clusters $C_n$ and on the reinforcing mechanism \citep[see][for a deep discussion]{lijoi2007controlling, favaro2013mcmc, argiento2016blocked}. 
We mention that both have an increasing effect on the probability of observing a new cluster and the prior (and posterior) number of clusters.
Interestingly, $\sigma$ also enters the expression of the weights of existing clusters and, as observed before, reduces the probability of clusters with few elements. We refer to this double effect of $\sigma$ as the  ``trade-off" between the number of clusters and reinforcement. In particular, for $(\kappa, \sigma)$ we adopted a discrete prior on a $10\times 10$ grid  in $(0, 15)\times(0.0, 0.6)$, such that the marginal distribution are discrete approximation of $\kappa \sim Gamma(2,1)$ and $\sigma \sim Beta(5,23)$, respectively.
Extending the work by \cite{muller2011product}, we aim at obtaining a prior for the random partition that encourages two subjects to co-cluster when they have similar covariates, i.e., predictive biomarkers. In particular, the prior on the random partition is defined perturbing the cohesion function of a product partition model in equation \eqref{eq:prodpart} via a similarity function $g$ inducing the desired dependence on covariates.  More in detail, the \emph{similarity} function $g$ is a non-negative function that depends on the covariates associated with subjects in each cluster. Let $\bm x_i$ denote the covariates for the $i-$th unit, while $\bm x_{j}^{\star} = (\bm x_i, i\in S_j)$ represents the covariates arranged by cluster. 
The product partition distribution with covariates is 
\begin{equation}
	\label{eq:ppmx}
	p(\mathcal{P}_n)\propto V_{n, C_n}\prod_{j=1}^{C_n}\rho(n_j)g(\bm x_{j}^{\star}).
\end{equation}

The choice of the similarity function is of paramount importance for our modeling. It measures the homogeneity of covariates arranged by clusters, and thus, the more the covariates take similar values, the larger the value of $g$ must be. The default choice, proposed by \cite{muller2011product}, defines $g$ as the marginal probability of an auxiliary Bayesian model. Several alternatives can be taken \citep[see for example][]{page2018calibrating, argiento2022clustering}, since the only requirement for $g$ is to be a symmetric non-negative function. We implement the ``Double Dipper'' similarity function because it has been shown to work well both in settings with a large number of covariates and in settings where prediction is the main inferential goal \citep{page2016spatial,page2018calibrating}: 
\begin{equation}
	\label{eq:simfundd}
	g(\bm x_{j}^{\star})=\prod_{q=1}^{Q}\int \prod_{i\in S_j}p(x_{iq}|\bm\xi_{j}^{\star})p(\bm\xi_{j}^{\star}|\bm x_{jq}^{\star})\mathrm{d} \bm\xi_{j}^{\star},
\end{equation}
with 
$p(\bm\xi_{j}^{\star}|\bm x_{jq}^{\star})\propto \prod_{i\in S_j}p(x_{iq}|\bm\xi_{j}^{\star})p(\bm\xi_{j}^{\star})$. This structure
is not due to any probabilistic properties since the covariates are not considered random, but it measures the similarity of the covariates in cluster $S_j$. The name comes from the fact that the covariates are used twice and correspond to the $\bm x_{j}^{\star}$'s posterior predictive. 
The model in equation \eqref{eq:simfundd} is completed by assuming $p(\cdot|\bm\xi_{j}^{\star})=N(\cdot|m_{j}^{\star}, v_{j}^{\star})$, where $N(\cdot|m, v)$ is a Gaussian density with mean $m$ and variance $v$, 
and $p(\bm \xi_{j}^{\star})=p(m_{j}^{\star}, v_{j}^{\star}) = NIG(m_{j}^{\star}, v_{j}^{\star}|m_0, k_0, v_0, n_0)$ is the Normal-Inverse-Gamma density function. The resulting similarity function can model scenarios with heterogeneous within-cluster variability. 
We follow \cite{page2018calibrating} and set the parameters of the Normal-Inverse-Gamma density to the default values $m_0 = 0, k_0 = 1.0, v_0 = 1.0, n_0 = 2$; 
since there is no notion of the $\bm x_i$ being random, parameters $\bm \xi^{\star}_{j}$ are not updated. 
Approaches based on covariate-dependent random partition perform well if the clustering is not completely driven by covariates. 
As the number of covariates increases, similarity functions tend to overwhelm the information provided by the response, completely driving the clustering process. 
To counteract this behavior, we calibrate the influence of covariates on clustering. To this end, with an abuse of notation, $g$ in equation \eqref{eq:ppmx} is taken to be $g^{}(\bm x_{j}^{\star}):=g(\bm x_{j}^{\star})^{1/\sqrt{Q}}$, namely a small variation of the \emph{coarsened similarity} function by \cite{page2018calibrating}. The impact of the cohesion and similarity functions on the number of clusters is evaluated in a simulation study reported in Supplementary Material B; in summary, this simulation study demonstrates the effectiveness of the NGGP in controlling the prior mass allocated to different partitions through the reinforcement mechanism induced by $\sigma$. Additionally, we observe that the covariates included in the prior effectively drive the clustering process, as desired.
\section{Posterior Inference and Treatment Selection}
\label{sec:ts}
We implement an MCMC algorithm to simulate from the posterior distribution of the parameters of interest. The core part of the MCMC algorithm is the update of cluster membership; the computation associated with the joint law of $(\mathcal{P}_{n^a}^a, \bm\eta_{j}^{a\star})$ is based on \cite{neal2000markov}'s Algorithm 8 with a reuse strategy \citep{favaro2013mcmc}. Conditional on the updated cluster labels, all the remaining parameters are easily updated with Gibbs sampler or Metropolis-Hastings steps. Details on posterior inference are given in Supplementary Material C.
To perform treatment selection for a new untreated patient, we need to predict the treatment outcome under each competing treatment $\tilde{y}^{a}$, for $a=1, \dots, T$. 
Given the observed responses $\bm y^a$ for the $n^a$ patients previously treated with therapy $a$,  the predictive probability of response level $k$ under treatment $a$ is $p(\tilde{y}^a=k\mid \bm y^a, \bm z^a, \bm x^a, \tilde{\bm z}, \tilde{\bm x})$, where $\tilde{\bm z}$ and $\tilde{\bm x}$ denote the new patient's biomarkers. 
%
%
%
We establish utility weights that turn a multinomial setting into a one-dimensional selection criterion considering the relative importance of each level of the ordinal response. 
Let $\bm{\omega}=(\omega_1, \dots, \omega_K)^\top$ be a $K-$dimensional vector denoting the utility assigned to tumor response levels. We can then compute the median predictive utility for a new patient treated with treatment $a$ as $\tilde{\varphi}(a, \bm\omega, \mu_{0.5})=\sum_{k=1}^{K}\omega_k \mu_{0.5}(\tilde{y}^a=k\mid \bm y^a, \bm z^a, \bm x^a, \tilde{\bm z}, \tilde{\bm x})$, 
where $\mu_{0.5}(\cdot)$ denotes the median of the posterior predictive distribution; see Supplementary Material C.3 for more details. 
The approach under consideration may not be suitable for complex settings where treatment selection depends on assessing multiple endpoints. \cite{lee202utility} suggest an alternative approach that involves eliciting a utility function dependent on both covariates and response to account for two endpoints.
Finally, an untreated patient will be assigned to the treatment with the highest predicted utility. 
%
%
%
\section{Simulation Study}
\label{sec:ss}
We carried out a comparative study on simulated data to evaluate the performance of our method. We compare the proposed integrative model for personalized treatment selection (t-ppmx) with the two-step predictive model proposed by \cite{ma2019bayesian} using, for the first step,  three alternative clustering procedures, namely K-means (km-bp), Partitioning Around Medioids (pam-bp) and Hierarchical Clustering (hc-bp). T-ppmx is also compared with a selection rule based on Dirichlet-multinomial (DM) regression model. We assume a horseshoe prior on the regression coefficients for a fair comparison. In the DM regression model, we included the main effects of prognostic and predictive biomarkers and all the interaction terms between predictive covariates and the treatment. 
We fit the models on the $n_{train}$ patients of the training set, and we evaluate the predictive performance on $n_{test}$ patients in the test set. 
The approach proposed by \cite{ma2019bayesian} employs the Consensus Clustering method \citep{monti2003consensus}, which determines clustering for a specified number of clusters $C$. Since the number of groups is unknown, $C$ must be selected using leave-one-out cross-validation for each simulated patient. Note that this is not true for our approach, and our method does not need to perform this extra step. 
We generate simulated datasets closely following the strategy devised in \cite{ma2019bayesian}, i.e., we do not employ our model as the generative mechanism. The patients are assigned to $T=2$ treatments, and $K = 3$ levels of the response variable are considered. Since the observed treatment endpoints were unavailable, the treatment response and the optimal treatment for each simulated patient were generated. In particular, we set $\bm\omega=(0, 40, 100)^\top$ to make the ordinal response reflect the clinical importance of each level; additional details on the data generating mechanism and on the weights elicitation are provided in Supplementary Material D and A, respectively. 
\subsection{Performance evaluation}
\label{sec:perfev}
Prediction performances are compared in terms of the following metrics:
\begin{itemize}
    \item[(i)] $MOT$: it counts the number of patients misassigned to their optimal treatment;
    \item[(ii)] $\%\Delta MTU$: it measures the relative gain in treatment utility;
    \item[(iii)] $NPC$: it counts the number of patients whose outcome has been correctly predicted.
\end{itemize}
The true optimal treatment for each simulated patient is available since has been determined as a result of the generating mechanism. 
$MOT$ represents a first measure to compare the methods, its interpretation is straightforward, and lower values are associated with better selection rules. 
Nonetheless, the extent to which a particular treatment is beneficial for each patient is heterogeneous, and the improvement offered by a therapy varies from patient to patient. 
The relative gain in Treatment Utility \citep{ma2016bayesian}, $\%\Delta MTU$, measures 
the overall benefit ensured by a treatment selection rule in the case of $T=2$ competing treatments. In particular, $\%\Delta MTU$ is bounded above by $1$ when it always recommends the optimal treatment, and $\%\Delta MTU=-1$ when it fails to select the optimal therapy for all the patients. More details on this measure can be found in Supplementary Material E. 
The $NPC$ metric represents the number of patients whose outcome is correctly predicted. 
\subsection{Simulation scenarios and results}
\label{sec:simres}
Methods are compared on scenarios of increasing complexity. We construct Scenarios 1a and 1b following the generating mechanism described in Supplementary Material D, using $2$ prognostic and $25$ and $50$ predictive biomarkers, respectively. 
Other scenarios emulate the pronounced heterogeneity that genomics data feature. Namely, the predictive covariates employed to generate the response differ in the train and the test set since they overlap only to some extent. The pairs of scenarios (2a, 2b) and (3a, 3b) match (1a, 1b) in the number of predictive covariates, but predictive markers employed to generate the response in the train and test set overlap at $90\%$ and $80\%$, respectively. Train and test sets consist of $124$ and $28$ observations, respectively. Scenarios with $25$ covariates are labeled with ``a'', while those with $50$ covariates with ``b''. Hyperparameters are set to the default values given in the previous sections; sensitivity to these settings is studied in Supplementary Material A. We run the algorithm for $12,000$ iterations, with a burn-in period of $2,000$ iterations; chains were thinned, and we kept every $5-$th sampled value. We fit the model on $50$ replicated datasets. 
Reported values are averaged over the replicated datasets, with standard deviations in parentheses.

\begin{table}
\caption{Predictive performance: mean across $50$ replicated datasets, standard deviations are in parentheses. In each scenario and for each index, the best performance is in bold.}
\label{tab:scen}
\begin{tabular*}{\columnwidth}{@{}l@{\extracolsep{\fill}}|c@{\extracolsep{\fill}}c@{\extracolsep{\fill}}c@{\extracolsep{\fill}}|c@{\extracolsep{\fill}}c@{\extracolsep{\fill}}c@{\extracolsep{\fill}}c@{}}
  \hline\hline
 & \multicolumn{3}{c}{\textbf{Scenario 1a}} & \multicolumn{3}{c}{\textbf{Scenario 1b}} \\
& $MOT$ & $\%\Delta MTU$ & $NPC$ & $MOT$ & $\%\Delta MTU$ & $NPC$\\
  \hline
\multirow{2}{*}{pam-bp} & 14.0600 & 0.0192 & 10.2600 & 14.2400 & -0.0106 & 13.2000 \\ 
                        & (3.2351) & (0.3447) & (1.9672) & (2.9593) & (0.3429) & (2.2039) \\[.05cm]
\multirow{2}{*}{km-bp}  & 13.4200 & 0.1130 & 11.4000 & 13.4000 & 0.0750 & 13.9600 \\ 
                        & (2.8074) & (0.3038) & (2.5314) & (2.6108) & (0.3076) & (2.2584) \\[.05cm]
\multirow{2}{*}{hc-bp}  & 12.8600 & 0.1520 & 12.0200 & 12.4400 & 0.1418 & 12.6800 \\ 
                        & (3.1429) & (0.3642) & (2.8961) & (3.1374) & (0.3403) & (2.7807) \\[.05cm]
\multirow{2}{*}{dm-int} & 12.6000 & 0.1756 & 13.8200 & 13.2800 & 0.0740 & 12.9600 \\ 
                        & (3.4934) & (0.3536) & (2.9877) & (3.7310) & (0.3851) & (3.0902) \\[.05cm]
\multirow{2}{*}{t-ppmx} & \bf{10.0000} & \bf{0.3933} & \bf{15.1600} & \bf{10.7800} & \bf{0.3339} & \bf{14.4280} \\ 
                        & (3.2451) & (0.3080) & (2.2800) & (3.2968) & (0.3362) & (2.8646) \\
   \hline
& \multicolumn{3}{c}{\textbf{Scenario 2a}} & \multicolumn{3}{c}{\textbf{Scenario 2b}} \\
& $MOT$ & $\%\Delta MTU$ & $NPC$ & $MOT$ & $\%\Delta MTU$ & $NPC$\\
  \hline
\multirow{2}{*}{pam-bp} & 14.1600 & 0.0145 & 10.1600 & 14.2000 & -0.0068 & 13.2200 \\ 
                        & (3.2474) & (0.3405) & (2.2439) & (2.9966) & (0.3487) & (2.2341) \\ [.05cm]
\multirow{2}{*}{km-bp}  & 13.3600 & 0.1136 & 12.4800 & 13.5200 & 0.0689 & 13.7800 \\ 
                        & (2.9190) & (0.3082) & (2.7198) & (2.5414) & (0.3037) & (2.2883) \\ [.05cm]
\multirow{2}{*}{hc-bp}  & 12.9600 & 0.1223 & 11.5400 & 12.4400 & 0.1430 & 12.7600 \\ 
                        & (3.5165) & (0.3846) & (11.54) & (3.1112) & (0.3344) & (2.7372) \\ [.05cm]
 \multirow{2}{*}{dm-int}& 12.9600 & 0.1223 & 11.5400 & 13.0400 & 0.1021 & 13.0200 \\ 
                        & (3.5165) & (0.3847) & (2.8082) & (3.5798) & (0.3627) & (2.9657) \\[.05cm]
\multirow{2}{*}{t-ppmx} & \bf{10.6200} & \bf{0.3578} & \bf{15.3800} & \bf{10.6000} & \bf{0.3497} & \bf{14.4400} \\
                        & (3.3313) & (0.3347) & (2.5446)  & (3.1880) & (0.3269) & (2.8224)  \\
   \hline
& \multicolumn{3}{c}{\textbf{Scenario 3a}} & \multicolumn{3}{c}{\textbf{Scenario 3b}} \\
& $MOT$ & $\%\Delta MTU$ & $NPC$ & $MOT$ & $\%\Delta MTU$ & $NPC$\\
  \hline
\multirow{2}{*}{pam-bp} & 13.9800 & 0.0275 & 11.8600 & 14.5000 & -0.0635 & 14.1400 \\ 
                        & (3.3654) & (0.3469) & (2.6955) & (2.9433) & (0.3310) & (2.8856) \\ [.05cm]
\multirow{2}{*}{km-bp}  & 13.3600 & 0.1159 & 11.6000 & 13.7000 & 0.0405 & 13.8000 \\ 
                        & (2.8909) & (0.3055) & (2.6954) & (2.9014) & (0.3363) & (2.5873) \\[.05cm] 
\multirow{2}{*}{hc-bp}  & 12.6600 & 0.1621 & 11.5000 & 12.9800 & 0.0859 & 12.3800 \\ 
                        & (3.2740) & (0.3684) & (2.8158) & (3.3715) & (0.3647) & (2.3724) \\ [.05cm]
\multirow{2}{*}{dm-int} & 12.7400 & 0.1616 & 13.9000 & 12.8400 & 0.0957 & 13.6000 \\ 
                        & (3.7461) & (0.3747) & (3.1445) & (3.1646) & (0.3548) & (2.9207) \\[.05cm]
\multirow{2}{*}{t-ppmx} & \bf{10.2600} & \bf{0.3610} & \bf{15.0400} & \bf{10.8600} & \bf{0.3244} & \bf{14.8400} \\ 
                        & (3.6411) & (0.3352) & (2.4320) & (3.1234) & (0.3281) & (2.6677) \\
   \hline
\end{tabular*}
\end{table}
Overall, t-ppmx outperforms all competing methods (Table \ref{tab:scen}). 
This result can be attributed to the ability of the covariate-dependent random partition to reach significant clustering arrangements. 
Among the two-stage methods, pam-bp consistently exhibits inferior performance compared to other methods, at least considering MOT and MTU. While km-bp and hc-bp produce comparable results, hc-bp demonstrates superior performance, especially when the number of covariates is large. However, km-bp exhibits greater robustness to increasing heterogeneity, at least for moderate numbers of covariates, and deteriorates less compared to hc-bp. Dm-int and hc-bp exhibit similar performance in scenarios with a moderate number of covariates. However, as the number of covariates increases, dm-int demonstrates greater robustness. Conversely, hc-bp shows superior robustness in scenarios with considerable heterogeneity.
Interestingly, t-ppmx is robust with respect to increasing heterogeneity. 
It is probably due to the integrated prediction mechanism, which fully accounts for the uncertainty in the clustering; note that, for the proposed method, optimal treatment misassignment often pertains to patients with similar utility across treatments. Our simulation study suggests that t-ppmx should be preferred over two-step methods. 
In this simulation study, no measure of clustering accuracy has been produced since the generative mechanism for synthetic data 
implies that no ``true'' clustering exists. To evaluate the clustering performances of t-ppmx, we design three scenarios where covariates have a known intrinsic data structure, and clustering explicitly depends on covariates. This enables us to compare PPMx's posterior partition with the true one. 
In Supplementary Material B we compare the methods on alternative scenarios to examine their robustness across different generating processes, and we also assess the impact of misspecification of prognostic and predictive covariates on t-ppmx's clustering and predictive performances. 
\section{Case Study of Low-grade Glioma}
Glioma is the most frequent brain tumor: it makes up approximately 30\% of all brain and central nervous system tumors and 80\% of all malignant brain tumors \citep{goodenberger2012genetics}. Gliomas are classified as grades I to IV based on histological criteria established by the World Health Organization (WHO). Grade I tumors are generally circumscribed benign tumors with favorable prognoses, while grades II-IV comprise more aggressive tumors (diffuse gliomas). Grade II and grade III gliomas are usually referred to as low-grade glioma (LGG), which may eventually progress to grade IV, high-grade glioma. 
Most LGG patients undergo resection and then receive radiotherapy and/or chemotherapy. Nonetheless, these standard procedures have proved to be largely inadequate \citep{claus2015survival}. LGG exhibits significant molecular heterogeneity
, and many research efforts are now devoted to developing precision medicine for these patients
\citep{olar2015molecular, ius2018nf}.
We apply our method to the dataset analyzed in \cite{ma2019bayesian}, where clinical data and protein expression of patients affected by lower-grade glioma are collected from the TCGA data portal (\url{https://portal.gdc.cancer.gov/}, accessed August 31, 2023).
Publicly available data underwent an accurate preprocessing, thoroughly documented in \cite{ma2019bayesian}, and summarized in Supplementary Material F. The resulting LGG dataset comprises patients that received standard and advanced treatments. A treatment qualifies as advanced if it includes targeted therapies or radiotherapy. Each group comprises 79 patients balanced in the covariates to account for potential selection bias. 
Following \cite{ma2019bayesian}, we defined the tumor response for the LGG dataset using the RECIST criteria (\url{http://www.recist.com/}, accessed August 31, 2023). 
In our analysis, tumor response is formulated in three ordinal levels: progressive disease (PD), partial response/stable disease (PS), and complete response (CR). 
Utility weights for treatment selection for ordinal outcomes are elicited. Namely, $\bm \omega = (0, 40, 100)^\top$ to make the ordinal response reflect the clinical importance of each level \citep{ma2016bayesian}. 
We evaluate the robustness of our method to weight elicitation in Supplementary Material I.  
Finally, we analyze the same 23 predictive and 2 prognostic protein expressions considered in \cite{ma2019bayesian}. See Supplementary Material F for more details, including the list of predictive and prognostic proteins.
TCGA data do not provide the true optimal treatment, and only the $NPC$ measure, among those discussed in Section \ref{sec:perfev}, can be used. 
We employ an empirical summary measure \citep[ESM,][]{song2004evaluating} to evaluate the relative increase in the population response rate attributable to a treatment allocation method compared to random allocation. Let $Y$ be the binary outcome variables, taking $0$ for non-respondents or $1$ for respondent patients. We define the treatment contrast as $\Delta(\bm X, \bm Z) = Pr(Y=1|A=2, \bm X, \bm Z)-Pr(Y=1|A=1, \bm X, \bm Z)$, where $A=\{1,2\}$ denote the non-targeted and targeted treatment, respectively. Indicating with $Pr(Y=1|A_r)$ the probability of being a respondent under a randomized treatment assignment, we obtain the relative increase in the population response rate under a personalized treatment selection rule as: 
$
ESM= 
\{Pr(Y=1|A=2, \Delta(\bm X, \bm Z)>0)\times Pr(\Delta(\bm X, \bm Z)>0)+ 
Pr(Y=1|A=1, \Delta(\bm X, \bm Z)<0)\times Pr(\Delta(\bm X, \bm Z)<0)\}-Pr(Y=1|A_r),
$
see Supplementary Material E for more details.
Note that we based this summary measure on only two response categories, responders (CR) and non-responders (PD + PS), whereas we used all three levels of the ordinal outcome in the data analysis and to implement personalized treatment selection. 

\subsection{Results}
\label{sec:results}
In this section, we applied the proposed method to the LGG dataset alongside the approach proposed by \cite{ma2019bayesian}. Table \ref{tab:predlgg} reports NPC and ESM summary measures computed from assignments obtained using a 10-fold cross-validation strategy. We run the algorithm for $12,000$ iterations, with a burn-in period of $2,000$ iterations; chains were thinned, and we kept every $5-$th sampled value. 
We report MCMC diagnostic checks in Supplementary Material F. 
\begin{table}[ht]
\centering
\caption{Predictive performance: metrics are obtained gathering 10-fold cross-validation results. For each index, the best performance is in bold.}
\begin{tabular}{@{}l@{\extracolsep{\fill}}|c@{\extracolsep{\fill}}c@{\extracolsep{\fill}}c@{}}
  \hline
  & NPC & ESM \\
  \hline
  pam-bp & 48 & 0.0553 \\
  km-bp & 45 & 0.0384 \\
  hc-bp & 48 & 0.0285 \\
  dm-int & 64 & 0.0746 \\
  t-ppmx & \bf{69} & \bf{0.1008} \\
  \hline
  \end{tabular}
\label{tab:predlgg}
\end{table}
The proposed t-ppmx outperforms competing methods both in terms of NPC and ESM. These results are consistent with those obtained in our simulations studies, especially in scenarios featuring significant heterogeneity and a moderate number of predictive covariates (Scenarios 2a and 3a). In particular, t-ppmx attains an ESM of $0.1008$, while ESM for pam-bp is $0.0553$ among two-stage procedures.
Patients show pronounced heterogeneity, particularly those assigned to Treatment 2. The absence of a sharp separation between clusters demonstrates a significant uncertainty in the clustering. Patients assigned to Treatment 1 form more homogeneous clusters, but the low probability of co-clustering still indicates a large variability in clusters' assignments. 
\subsection{Cluster Analysis}
\label{sec:cluana}
Here, we want to investigate the composition of the clusters identified to characterize the profiles of co-clustered patients. 
\begin{figure}[h]
    \centering
    \includegraphics[width = .9\linewidth, keepaspectratio]{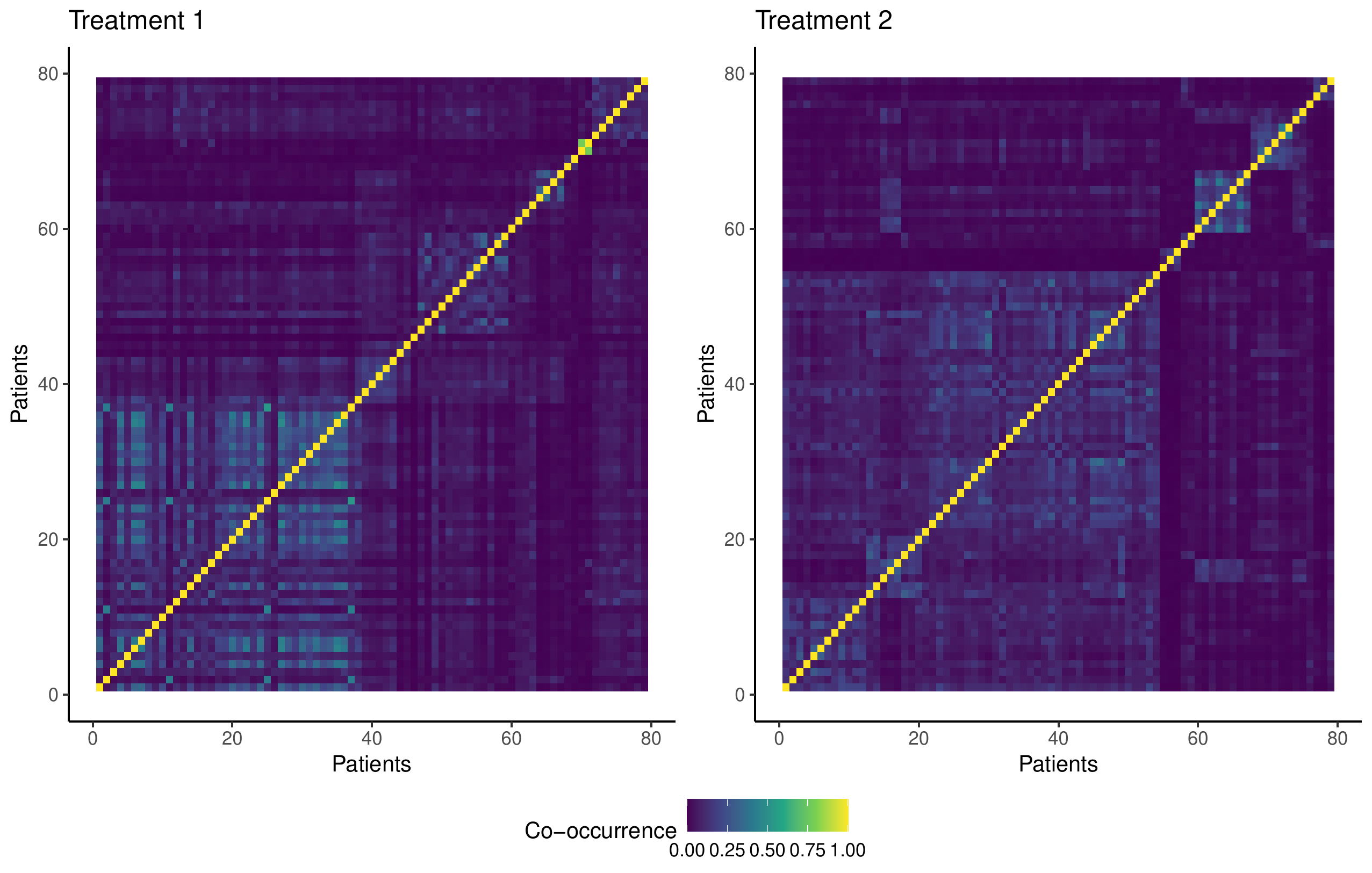}
    \caption{Heatmap of averaged co-occurrence matrix for patients that received Treatment 1 (left) and Treatment 2 (right). }
    \label{fig:avg_coocc}
\end{figure}
Following \cite{wade2018bayesian}, we use the variation of information loss function to estimate the optimal partition on the space of clusters. 
In particular, we obtain a partition of the 79 patients that received the \emph{standard treatment} (Treatment 1) into 10 groups ranging from 1 to 38. Similarly, patients that received the \emph{advanced treatment} (Treatment 2) are grouped into 10 clusters with cluster membership ranging from 1 to 34.
Figure \ref{fig:avg_coocc} reports the heatmap of the averaged co-occurrence matrices. 
We refer to T1G1,$\ldots$, T1G10 to denote the groups of patients treated with the standard treatment (left panel of Figure \ref{fig:avg_coocc}) and to T2G1,$\ldots$, T2G10 for the groups of patients that received the advanced treatment (right panel). 
Our PPMx model provides homogeneous clusters in terms of predictive covariates; 
indeed, it substantially reduces the within-group variance for each predictive covariate (see Figure 4 in Supplementary Material G). 
We deem clusters with less than 8 members residual clusters and exclude them from the following analysis.  
\begin{figure}[h]
    \centering
    \includegraphics[width = .9\linewidth, keepaspectratio]{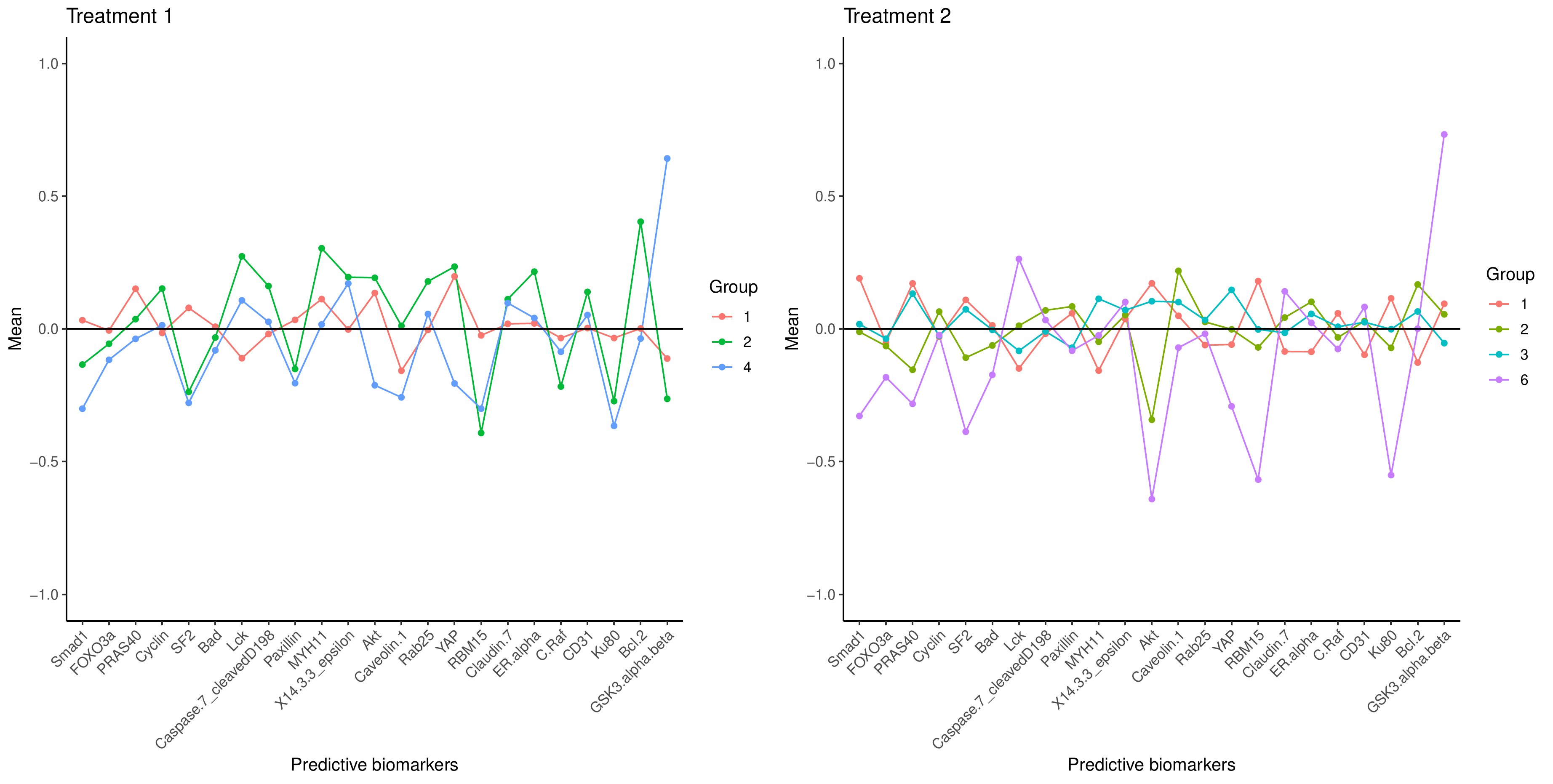}
    \caption{Group-specific mean of predictive biomarkers for patients that received Treatment 1 (left) and Treatment 2 (right). }
    \label{fig:mean_group}
\end{figure}
To characterize the groups, we consider the cluster-specific mean for predictive biomarkers. Figure \ref{fig:mean_group} shows that cluster-specific means in T1G2, T1G4, and T2G6 strongly depart from the population value. Moreover, T1G2 and T1G4 feature the under-expression of a mutual set of proteins, namely SF2, RB15, and KU80, still presenting opposite trends in the expression of AKT, YAP, BC2, and GSK3. 
Cluster-specific means in T2G1, T2G2, and T2G3 are really close for almost all the proteins. Noticeably, T2G3 features a sharp under-expression of AKT with respect to the mean values expressed in T2G1 and T2G2. 
Finally, patients in T2G6 show under-expression of SF2, AKT, RBM15, RBM15, and KU80, in addition to the over-expression of GSK3. It is important to note that both T1G4 and T2G6 exhibit similar protein expression patterns. Specifically, both groups displayed under-expression of SF2, AKT, RBM, and KU80, as well as over-expression of GSK3. The role of these proteins in tumor progression is not entirely understood. Nonetheless, most of these proteins have been implicated in gliomas' oncogenesis and developmental process \citep{mills2011emerging, li2020downregulation}.
A better characterization of the groups of interest can be achieved by evaluating the cluster-specific response probability. 
To obtain meaningful cluster-specific parameters we consider the \emph{a posteriori} estimated clustering $\hat{\mathcal{P}}^{a}_{n^a}$ as fixed, and --conditional on it-- we obtain the \emph{a posteriori} distribution of $\bm\pi^{a\star}_j$s. 
Response probabilities are summarized by the posterior distributions $\bm\pi_{j}^{a\star}\mid \hat{\mathcal{P}}^{a}_{n^a}, \bm x^a$ displayed in Figure \ref{fig:tern}. 
  \begin{figure}[hbt!]
    \centering
    \includegraphics[width = .9\linewidth, keepaspectratio]{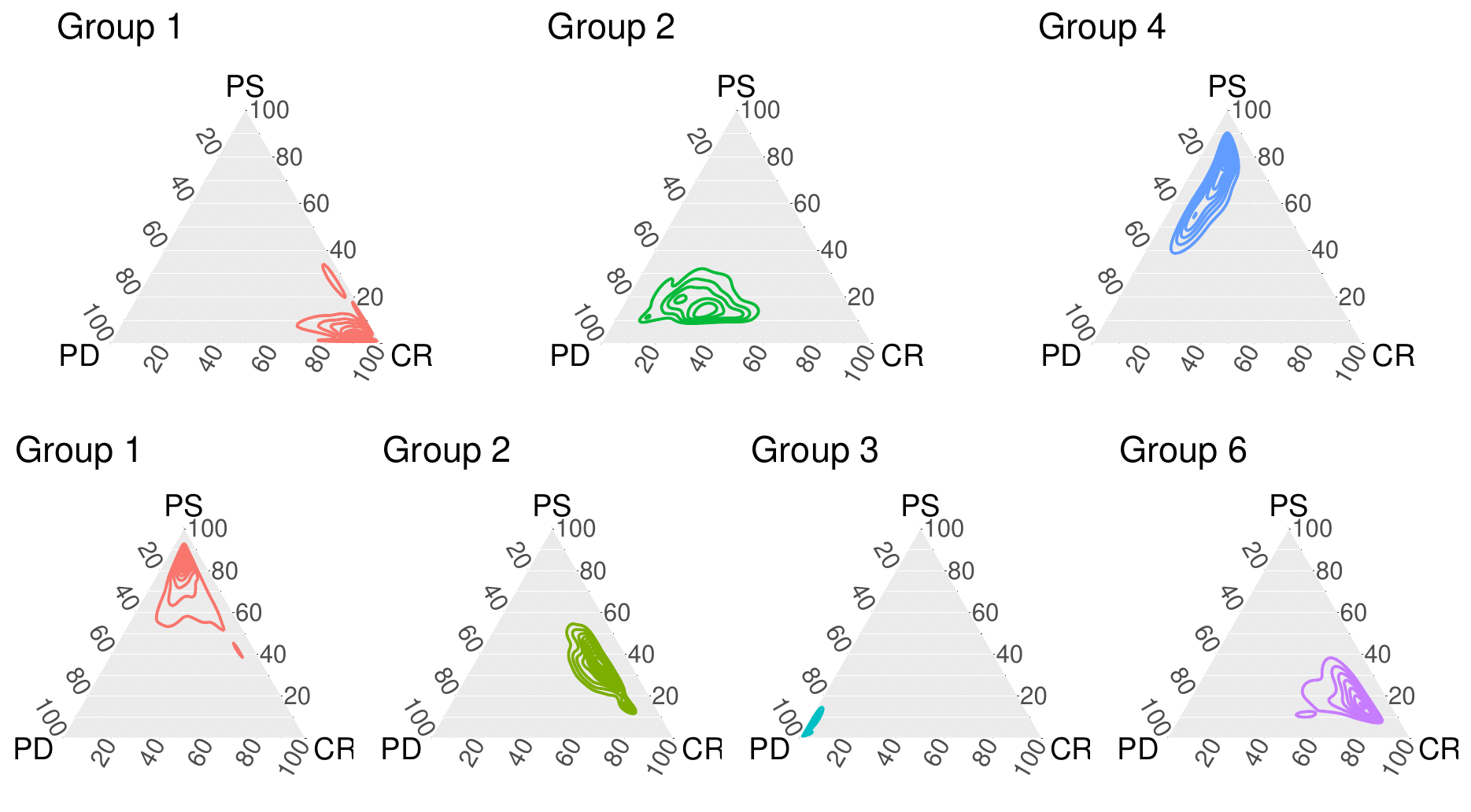}
    \caption{Ternary plot of the posterior density of group-specific response probabilities for patients that received Treatment 1 (first row) and Treatment 2 (second row). }
    \label{fig:tern}
\end{figure}
Figure \ref{fig:tern} displays the response probabilities for patients that underwent the standard treatment (first row). Patients in T1G1 are those that most benefit from the standard treatment, as the posterior distribution of the response probability is concentrated toward the CR vertex. On the other hand, patients in T1G4 and T1G2 are more likely to experience a partial or non-response to the treatment.
Response probabilities for patients that received Treatment 2 (Figure \ref{fig:tern}, second row) clearly characterize these clusters of patients, too. Notably, the posterior densities exhibit an evident skewness towards the vertices of the ternary plot. 
If we consider T2G1, T2G2, and T2G3, our model successfully uses predictive biomarkers along with the response to the treatment to cluster patients. In fact, t-ppmx is able to distinguish between patients who may be similar in terms of their covariates (see \ref{fig:mean_group}, right panel) but have different responses to treatment, which is a common phenomenon in cancer genetics. 
Nonetheless, it is also important to notice that discrimination of respondents may be accomplished based on AKT protein underexpression.
Among the Treatment 2 clusters, T2G6 is characterized by unique patterns in terms of posterior probabilities and cluster-specific means. 
On a comparative analysis, it is noteworthy that T2G6 and T1G4 exhibit a shared set of under/up-regulated proteins, namely SF2, AKT, RBM15, and KU80 are under-expressed, and GSK3 is over-expressed. Consequently, T1G4 and T2G6 subjects can be regarded as individuals with closely-related genetic profiles that underwent different treatment modalities. Intriguingly, these patients have shown limited response to the standard procedure (Treatment 1), whereas they appear to be responsive to the advanced treatment (Treatment 2).
Interpretation of the within-cluster expression levels needs particular care.
Interactions among the proteins and the relationships between proteins' expression and tumor progression are highly complex. Nonetheless, our analysis provides a proof of concept that the proposed method can empirically identify subgroups in heterogeneous populations. Noteworthy, the proposed method quantifies the group-specific deviation from a population baseline accounting for the variability in the clustering and, in contrast to what is typically done with regression models, does not require us to prespecify the functional form of the association between predictive covariates and the outcome variable.

\section{Discussion}
\label{sec:disc}
We have proposed a novel Bayesian approach that, given a set of predictive and prognostic biomarkers, suggests the best-suited treatment for each patient. The model clusters patients into homogeneous groups with respect to their predictive markers, separately for each treatment. Cluster-level effects adjust the baseline probability of response to treatment obtained by prognostic factors. As a key innovative feature of the proposed approach, model-based clustering and treatment assignment are jointly estimated from the data, that is, treatment selection fully accounts for patients' heterogeneity. 
Simulation studies and the analysis of LGG data showed that the proposed method is well suited for predictions in scenarios of practical relevance, e.g., in the presence of considerable heterogeneity. Moreover, our approach leads to a precise characterization of the clusters of patients supported by the data, identifying the group of patients more likely to benefit from targeted treatments.
In its current version, the model is designed to be used after the biomarker discovery phase, i.e., after identifying relevant prognostic and predictive biomarkers. This limitation could be addressed by adopting variable selection approaches in the Bayesian framework. Nevertheless, while the use of the latter methods is straightforward when selecting prognostic biomarkers entering the likelihood, variable selection methods for product partition models are part of our ongoing research \cite[see for instance][]{barcella2017comparative}. 
In this regard, we would like to highlight that, although assuming which are the prognostic and predictive biomarkers may be restrictive in certain scenarios, and we could recast the proposed model to accommodate this lack of knowledge, this assumption remains practically very relevant. In fact, the major drawback of a model that simultaneously performs biomarker discovery and treatment selection would be the absence of a confirmatory process. Biomarkers can lead to targeted therapy and serve as useful prognostic and predictive factors of clinical outcomes. Nonetheless, biomarkers need to be validated on a completely independent data set not used during development to serve these purposes. 
\section*{Acknowledgements}
We thank J. Ma for providing us with the companion \textsf{R} code of \cite{ma2019bayesian}. The first and third authors were partially supported by the ``Dipartimenti Eccellenti 2023-2027'' ministerial funds (Italy). 
All authors were partially supported by grant ``CLUstering: Bayesian Partition Models for Precise Medicine (CluB: PMx$^2$)'', funded by \emph{Fondo di Beneficienza di Intesa San Paolo} (Italy), the last author was partially supported by the Tuscany Health Ecosystem (THE) grant, funded by \emph{Ministero dell’Università e della Ricerca}.

\section*{Supporting Information}
The \textsf{R} code is also available on GitHub: \url{https://github.com/mattpedone/treatppmx}. Results and Figures can be reproduced using the scripts available at \url{https://github.com/mattpedone/Reproduce-tPPMx}.
\clearpage
%
%
\appendix
\setcounter{table}{0}
\renewcommand{\thetable}{A\arabic{table}}

\section{Sensitivity Analysis}
\subsection{Hyperparameter settings}
\label{app:sens}
Our predictive model involves some hyperparameters that require careful tuning. To this end, we investigate the sensitivity of the results to these values.
In particular, we construct the sensitivity study on Scenario 1a presented in Section 7.3, which considers $25$ predictive and $2$ prognostic biomarkers. Also, in Scenario 1a, we assign $152$ patients to $2$ competing treatments, and $K = 3$ levels of the ordinal response are assumed. We adopt a train and test strategy. 

For different specifications of parameters $\nu_0$ and $\bm \Lambda_{0}$, we evaluate the model's performance in terms of prediction, goodness-of-fit, and clustering production. 
In particular, we assume that $\bm\Lambda_0$ is a diagonal matrix, and we evaluate the sensitivity of the results to their specification over the following grid of values:
\begin{enumerate*}
    \item[(i)] $\nu_0 = \{0.1, 1.0, 10.0\}$;
    \item[(ii)] $\Lambda_{0_{kk}}=\{0.1, 1.0, 10.0\}$;
\end{enumerate*}
for $k=1, \dots, K$. 

To assess treatment selection performance, we use the summary measures discussed in Section 7.2. We also report the log-pseudo-marginal-likelihood ($lpml$), a goodness-of-fit metric \citep{christensen2011bayesian} that accounts for model complexity. Finally, to take into consideration also the cluster arrangement produced, we report the variation of information \citep[$VI$,][]{wade2018bayesian}. 

We run the algorithm for $12,000$ iterations, with a burn-in period of $2,000$ iterations; chains are thinned, and we keep every $5-$th sampled value. The analysis of each configuration is replicated, and the results averaged over $50$ runs, with standard deviations in parentheses. Results are reported in Tables \ref{tab:senspre} and \ref{tab:sensclu}. In Tables \ref{tab:senspre} and \ref{tab:sensclu}, $\{\Lambda_{0_{kk}}\}$ denote the set of $K$ elements on the diagonal of $\bm \Lambda_0$. 
Since clustering is performed independently across treatments, results are reported separately for each treatment. In Table \ref{tab:sensclu}, T1 and T2 refer to Treatment 1 and Treatment 2, respectively.
We recall that the generative mechanism for synthetic data presented in Section 7.1 implies that no ``true'' clustering exists. Consequently, we can not establish the best performance in terms of clustering production. 

\begin{table}[ht]
\centering
{\begin{tabular}{ll|rr|rr|rr}
  \hline \hline    & & \multicolumn{2}{c}{$\nu_0 = 0.1$} & \multicolumn{2}{c}{$\nu_0 = 1.0$} & \multicolumn{2}{c}{$\nu_0 = 10.0$} \\
  \hline 
\multirow{3}{*}{$\{\Lambda_{0_{kk}}\} = 0.1$}  
 & $MOT$ & 10.4400 & (4.1264) & 10.4800 & (3.5181) & 10.9800 & (3.5021) \\ 
 & $\%\Delta MTU$ & 0.3631 & (0.4298) & 0.3507 & (0.3518) & 0.3359 & (0.3498) \\ 
 & $NPC$ & 14.9400 & (2.5826) & 14.7400 & (2.5299) & 15.5000 & (2.0727) \\ 
 & $lpml$ & -90.21869 & (16.5508) & -87.8940 & (9.2803) & -86.9153 & (11.2025) \\ 
 \hline
\multirow{3}{*}{$\{\Lambda_{0_{kk}}\} = 1.0$}  
 & $MOT$ & 10.7400 & (3.3124) & 10.1200 & (3.5722) & 10.2200 & (3.0257) \\ 
 & $\%\Delta MTU$ & 0.3391 & (0.3433) & 0.3946 & (0.3381) & 0.4121 & (0.2785) \\ 
 & $NPC$ & 15.1600 & (2.2620) & 15.2200 & (2.5737) & 15.4400 & (2.2054) \\ 
 & $lpml$ & -88.5130 & (14.5700) & -87.1011 & (12.8265) & -90.3171 & (13.2682) \\ 
\hline
\multirow{3}{*}{$\{\Lambda_{0_{kk}}\} = 10.0$}  
 & $MOT$ & 10.6800 & (3.1197) & 10.3300 & (3.4890) & 10.0000 & (3.2451) \\ 
 & $\%\Delta MTU$ & 0.3485 & (0.3071) & 0.3913 & (0.3455) & 0.3933 & (0.3080) \\ 
 & $NPC$ & 14.6800 & (2.3599) & 15.2000 & (2.7701) & 15.1600 & (2.2800) \\ 
 & $lpml$ & -88.4072 & (13.2897) & -86.4222 & (10.7379) & -86.4955 & (10.9973) \\ 
   \hline
\end{tabular}}
\caption{Predictive and goodness-of-fit performance: mean across $50$ replicated datasets, standard deviations are in parentheses. For each index the best performance is in bold.} 
\label{tab:senspre}
\end{table}

\begin{table}[ht]
\centering
\begin{tabular}{ll|rr|rr|rr}
  \hline \hline    & & \multicolumn{2}{c}{$\nu_0 = 0.1$} & \multicolumn{2}{c}{$\nu_0 = 1.0$} & \multicolumn{2}{c}{$\nu_0 = 10.0$} \\
\hline
& & T1 & T2 & T1 & T2 & T1 & T2 \\ 
  \hline
\multirow{2}{*}{$\{\Lambda_{0_{kk}}\} = 0.1$} 
 & \multirow{2}{*}{$VI$}   & 7.6000 & 8.0000 & 7.7000 & 8.000 & 7.2000 & 8.0000 \\ 
  &      & (0.8330) & (0.0000) & (0.5440) & (0.0000) & (1.0498) & (0.0000) \\ 
        \hline
 \multirow{2}{*}{$\{\Lambda_{0_{kk}}\} = 1.0$} 
 & \multirow{2}{*}{$VI$}    & 7.6200 & 7.9600 & 7.3800 & 8.0000 & 7.4200 & 7.9800 \\ 
 &       & (.6966) & (0.1979) & (0.8781) & (0.0000) & (0.8104) & (0.1414) \\ 
        \hline
\multirow{2}{*}{$\{\Lambda_{0_{kk}}\} = 10.0$} 
 & \multirow{2}{*}{$VI$}    & 7.4600 & 8.0000 & 7.5400 & 8.0000 & 7.4600 & 8.0000 \\ 
 &       & (0.8855) & (0.0000) & (0.9091) & (0.0000) & (0.9304) & (0.0000) \\ 
   \hline
\end{tabular}
\caption{Cluster production performance: mean across $50$ replicated datasets, standard deviations are in parentheses.}
\label{tab:sensclu}
\end{table}

The results are robust to different hyperparameter specifications, and we do not observe any sensitivity of the result to the values' specification. In the simulation study and the case study reported in the paper, we set $\{\Lambda_{0_{kk}}\} = 10.0$, and $\nu_0 = 10.0$. 

\subsection{PS weights elicitation}
Following \cite{ma2019bayesian}, the weight assigned to the partial response/stable disease (PS) was determined using a utility-based criterion, to take into consideration its clinical importance. 
Weights were given on a scale of 0 to 100, with 0 being the least favorable response level and 100 being the most favorable response level ($\omega_0 = 0$ and $\omega_{K} = 100$). The weight for the intermediate response level (PS) was determined based on its benefit in terms of long-term overall survival duration, estimated through a Cox regression model, as recommended by \cite{ma2016bayesian}. A landmark analysis was performed over a period of 120 days, given that a significant proportion of the observed responses occurred subsequent to the completion of two consecutive eight-week treatment cycles. Specifically, the estimated relative risk of $10-$year overall survival for response level PS was $2.46$ when adjusted for age, gender, tumor grade, and initial year of pathological diagnosis (IYPD), with complete response (CR) as the reference. As a result, the utility weight for PS was defined as $(1/2.46)100 \approx 41$. 
While the weights were not entirely arbitrary, we acknowledge that there may be alternative weighting schemes that could be considered, and to check the robustness of our method, we conducted a sensitivity study in which we varied the weight for PS, $\omega_{PS} = \{20, 40, 60, 80\}$, and re-analyzed the data. The results are displayed in Table \ref{tab:pswe}.
    
    \begin{table}[ht]
    \flushright
    \resizebox{.93\textwidth}{!}{%
        \begin{tabular}{c|cc|cc|cc|cc}
        \toprule
        \toprule
            & \multicolumn{2}{c}{$\omega_{PS} = 20$}  & \multicolumn{2}{c}{$\omega_{PS} = 40$}  & \multicolumn{2}{c}{$\omega_{PS} = 60$}  & \multicolumn{2}{c}{$\omega_{PS} = 80$} \\
        \midrule
        $MOT$ & 11.2200 & (2.2614) & 10.0000 & (3.2451) & 7.7600 & (5.3321)  & 6.1800 & (4.8051) \\ 
        $\%\Delta MTU$ & 0.3395  & (0.2441) & 0.3933  & (0.3080) & 0.5621 & (0.3732)  & 0.6542 & (0.3629) \\ 
        \bottomrule
        \end{tabular}
    }%
    \caption{Predictive performance: mean across $50$ replicated datasets, standard deviations are in parentheses.}
    \label{tab:pswe}
    \end{table}

    Our sensitivity analysis revealed that the treatment selection is sensitive to the weight elicited for PS. However, we have provided strong and valid justifications for our weight elicitation approach, which we consider to be conservative. In fact, we have observed that setting $\omega_{PS} = 0.40$ may have actually underestimated the true performance of our method, rather than overestimating it. Despite the sensitivity of our results to the weight elicitation method, we believe that our justifications provide a reasonable degree of confidence in the reliability of our findings.
    
\section{Additional Simulation Studies}
To further assess the performances of our proposed method, we conduct additional simulation studies. These simulations have multiple purposes: \begin{enumerate*} \item[(i)] to study the effect of the proposed cohesion and similarity functions on the number of induced clusters (Section \ref{sec:numclu}); \item[(ii)] to test the method's robustness under a different generating process  (Section \ref{ssec:lgm}); \item[(iii)] to evaluate the method's ability to recover true clustering structure in the data (Section \ref{ssec:kcs}); \item[(iv)] to assess the method's robustness to misspecification of prognostic/predictive biomarkers  (Section \ref{ssec:misspec}).\end{enumerate*}

\subsection{Induced prior distribution of the number of clusters}
\label{sec:numclu}
In this section, we empirically verify that our non-parametric prior has the features needed for our application. We argue that the Normalized Generalized Gamma process (NGGP) controls the prior mass allocated to different partitions through the reinforcement mechanism induced by $\sigma$. Moreover, we expect the incorporation of covariates in the prior to effectively drive the clustering process. These advantages can be appreciated by evaluating the induced prior distribution on the number $C_n$ of clusters. In order to assess the advantages offered by our covariate-dependent prior for random partitions, we consider a simple comparative example.
To this end, we compare the prior distribution on the number of clusters implied by different models for the random partition. Namely, subsequent case (v) coincides with equation 6 in the paper (i.e. the prior adopted in this paper), while experiments (i)-(iv) are relevant sub-cases of (v):

\begin{itemize}
    \item[(i)] $\rho(n_j)= \kappa(n_j-1)!$ and $g\equiv 1$, that yields to a product partition distribution coinciding with the eppf induced by a DP; we will refer to this as DP in the following;
    \item[(ii)] $\rho(n_j)= (1-\sigma)_{n_j}$ and $g\equiv 1$, that yields to a product partition distribution coinciding with the eppf induced by an NGGP; we will refer to this as NGGP in the following;
    \item[(iii)] $\rho(n_j)= \kappa(n_j-1)!$ and $g$ defined as $g(\bm x_{j}^{\star})=g(\bm x_{j}^{\star})^{1/\sqrt{Q}}$, that yields a product partition distribution with covariates with calibrated similarity and whose cohesion coincides with the eppf induced by a DP; we will refer to this as DP-sim in the following;
    \item[(iv)] $\rho(n_j)= (1-\sigma)_{n_j}$ and $g$ defined as in equation 7 in the paper, that yields a product partition distribution with covariates with non-calibrated similarity and whose cohesion coincides with the eppf induced by an NGGP; we will refer to this as NGGP-nocal in the following;
    \item[(v)] $\rho(n_j)= (1-\sigma)_{n_j}$ and and $g$ defined as $g(\bm x_{j}^{\star})=g(\bm x_{j}^{\star})^{1/\sqrt{Q}}$, that yields a product partition distribution with covariates with calibrated similarity and whose cohesion coincides with the eppf induced by a NGGP; we will refer to this as NGGP-sim in the following;
\end{itemize}
We fix $n = 50$ and consider the corresponding distributions of the number of components in the five above cases. We set hyperparameters of nonparametric priors such that the prior expected number of clusters, without the effect of the covariates, is $\mathbb{E}(C_{50})=25$; specifically, we set $\kappa = 19.2333$ for special cases (i) and (iii), and $\kappa = 0.7353, \sigma = 0.7353$ for (ii), (iv) and (v).

\begin{figure}
    \centering
    \includegraphics[scale=.4]{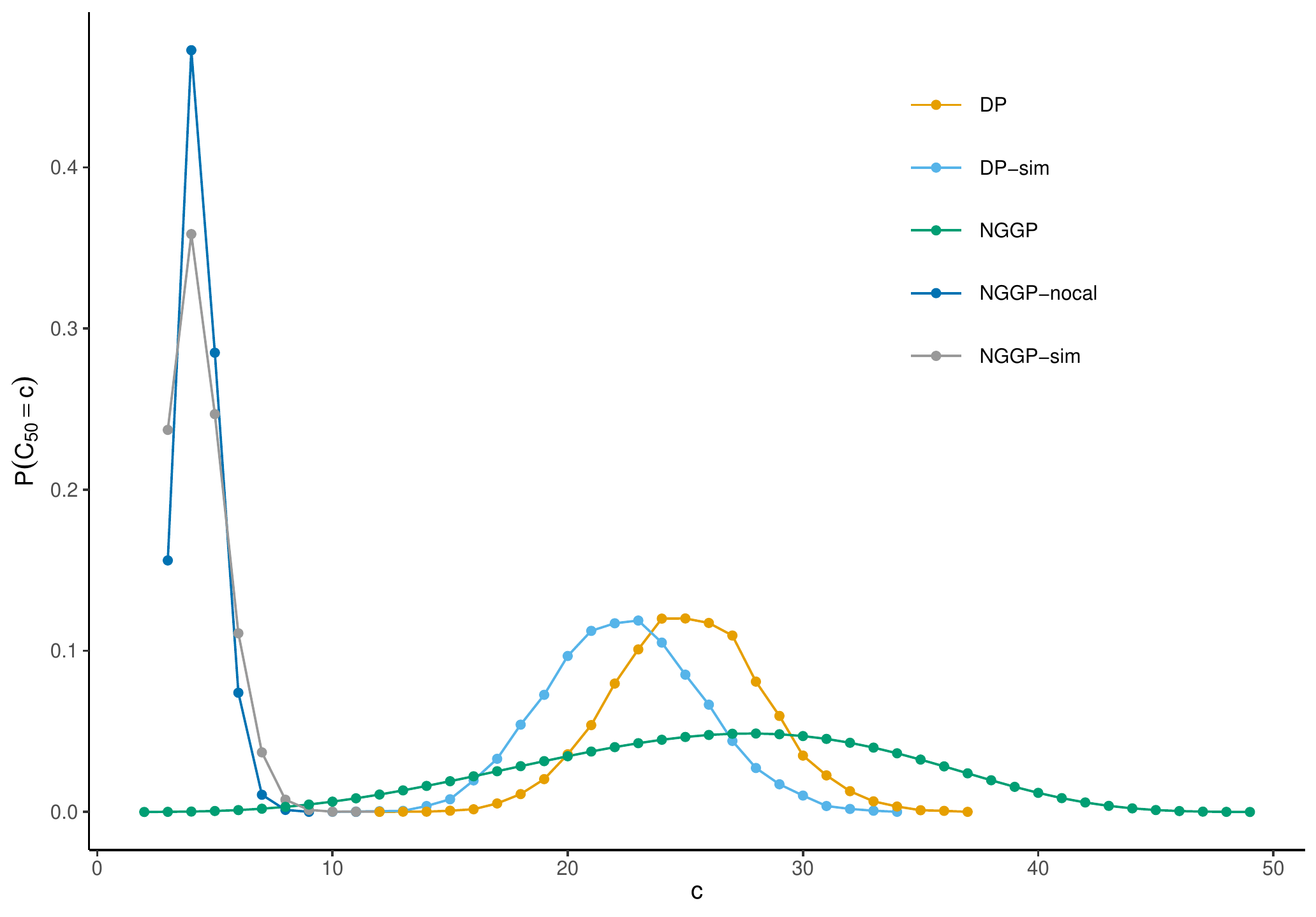}
    \caption{Prior distributions on the number of clusters corresponding to the 5 priors considered. 
    }
    \label{fig:pnc}
\end{figure}

Finally, we generate the covariates from a 3-component mixture of $5-$variate normal distributions such that
$$p(\bm x)=\sum_{j=1}^{3}\phi_j N_5(\bm\theta_j, \bm\Sigma),$$
where $\bm \phi =(0.2, 0.5, 0.3)^\top$, $\bm \theta_j$ are $5-$dimensional vectors such that $\bm\theta_1=-2.1\bm1$, $\bm\theta_2=0\bm1$ and $\bm\theta_3=2.3\bm1$, where $\bm1$ is the all-ones vector in $\mathbb{R}^5$. Finally $\bm\Sigma$ is $5\times 5$ diagonal covariance matrix such that $\bm\Sigma=diag(0.5, 0.5, 0.5, 0.5, 0.5)$.

As expected, NGGP results in a prior distribution of $C_{50}$ that is rather flat and exhibits a larger variability than the DP-induced distribution. In fact, due to the reinforcement mechanism induced by the $\sigma$ parameter, the NGGP prior gives \emph{a priori} support to a wider range of the number of clusters, still penalizing the number of singleton partitions (see Table \ref{tab:pnc}). This feature is particularly useful when little is known about the true number of clusters. The PPMx models include information from the covariates, and counteract the misspecification in the prior elicitation of the number of clusters. Nonetheless, when the DP cohesion is adopted, PPMx still does not differ much from the DP, exhibiting a distribution that supports more clusters than the truth. This is due to the rich-get-richer phenomenon, since a large portion of clusters ($53\%$) are singletons, as displayed in Table \ref{tab:pnc}. 

\begin{table}
\caption{Average number of clusters and proportion of singletons corresponding to the 5 priors considered.} 
\label{tab:pnc}
{\begin{tabular*}{\columnwidth}{@{}l@{\extracolsep{\fill}}c@{\extracolsep{\fill}}c
@{\extracolsep{\fill}}c@{\extracolsep{\fill}}c@{\extracolsep{\fill}}c@{\extracolsep{\fill}}c@{}}
  \hline \hline & DP & DP-sim & NGGP & NGGP-nocal & NGGP-sim \\ 
  \hline 
Av. \# clusters & 25.09 & 22.38 & 26.25 & 4.32 & 4.38 \\[.25pt]
  \% singletons & 55.82\% & 53.37\% & 34.45\% & 9.98\% & 19.39\% \\ 
   \hline
\end{tabular*}}
\bigskip
\end{table}

Finally, we focus on the PPMx models with NGGP as cohesion function. We consider both the case of calibrated and uncalibrated similarity functions. The implied distributions on the number of clusters give strong support to a moderate number of clusters, in both cases. 
Nonetheless, when the similarity is not calibrated, the PPMx implies a highly peaked distribution of $C_{50}$. This phenomenon is consistent with what is observed in \cite{page2018calibrating}, where they draw attention to the risk of similarities entirely driving the clustering process. Since this phenomenon is more pronounced for a larger number of covariates (as in our case study), the calibrated similarity is better suited for our application. \\

\noindent
In this simulation study we focused our attention on the prior distribution on the number $C_n$ of clusters induced by our prior for random partitions. 
Experiments (i) and (ii) represent special cases of our covariate-dependent prior, where the covariates do not guide the construction of the partition. DP and the class of Gibbs-type priors (to which NGGP belongs) have been widely studied in the bayesian nonparametric literature. Although we do not aim to provide a comprehensive study of these priors here, we find it interesting to highlight that our results are consistent with the theoretical results obtained by \citet{lijoi2007controlling} and \cite{de2013gibbs}. Furthermore, the following paragraph provides an opportunity to further explain the phenomena of ``rich-get-richer'' and reinforcement mechanisms mentioned earlier.

As we mention in the manuscript (see Section 4), when the Bayesian nonparametric prior is of  Gibbs-type, the cohesion assumes the analytical expression  $\rho(S_j)=(1-\sigma)_{n_j}$ with $\sigma<1$ and $(1-\sigma)_{n_j}$ being the rising factorials. 
It is evident that in this case, $\rho(S_j)$ is an increasing function of the cluster size $n_j$. So heavily populated clusters are more likely. This phenomenon is named the ``rich-get-richer'' behaviour of the clustering induced by a nonparametric prior. 
When $\sigma=0$, i.e., when the nonparametric prior is the Dirichlet process, the ``rich-get-richer'' behaviour is quite pronounced. As a consequence, frequently (see also Table \ref{tab:pnc}), the posterior cluster estimate is made by a few very populated clusters and many noisy clusters (i.e. clusters with few elements, or singletons). We refer to \cite{poux2021powered} for a detailed discussion on this.

As shown in our simulation study and discussed in \citet{lijoi2007controlling, favaro2013mcmc, argiento2016blocked}, to mitigate this behaviour, it is convenient to assume $\sigma>0$ by choosing a Normalized Generalized Gamma process as mixing distribution. In the following we briefly summarize the role of the parameter $\sigma$.

Following the notation adopted in the paper, the NGGP is indexed by $\sigma$, that controls the clustering, and $\kappa$, which plays the role of the mass parameter as in the Dirichlet process, which is recovered when $\sigma = 0$.
For detailed discussions on the effects of $\sigma$ and $\kappa$ on the prior (and posterior)  number of clusters  see \citet{lijoi2007controlling, favaro2013mcmc, argiento2016blocked}.  

We mention here that the integral defining the $V_{n,C_n}$ in equation (5) depends on $\kappa$ and $\sigma$. Further developing this expression \citep[see ][]{lijoi2007controlling, favaro2013mcmc} it is possible to show that: 
        
\begin{itemize}
    \item[(a)] when $\kappa$ and $\sigma$ increase, then the prior (and consequently the posterior) mean of $C_n$ increases;
    \item[(b)] large values of $\sigma$ imply heavy tail distribution on the prior for $C_n$.
\end{itemize} 
        
Moreover, it is possible to study the effect of $\kappa$ on the reinforcement mechanism. Indeed, when  $\sigma=0$, i.e. when considering the DP,  $\frac{V_{n+1,C_{n}+1}}{V_{n,C_n}}$ $\propto \kappa$ and  $\frac{V_{n+1,C_{n}}}{V_{n,C_n}}$ $\propto 1$.  Then, under the Dirichlet process, increasing $\kappa$ increases the probability of observing a new cluster. The same behaviour can be deduced when $\sigma>0$ using, for instance, the parameterization in \cite{favaro2013mcmc} (see their Section 3.1). 
        
Summing up, both parameters have an effect on the number of clusters and the reinforcing mechanism. Interestingly, $\sigma$ also enters the expression of the weights of existing clusters by reducing the probability of those with few elements. We refer to this double effect of $\sigma$ as the  ``trade-off'' between the number of clusters and reinforcement.

\subsection{Linear generating model}\label{ssec:lgm}
In this simulation study we use a mechanism to generate the response that can be regarded as a specific case of the generating process presented in Section \ref{sec:gtr}. 
We do not apply any transformation to the data entering the continuation-ratio logistic function that takes predictive covariates as argument. In particular, right hand side of \eqref{eq:crl1} reduces to $\alpha^{a}_{k}+\bm\phi^{a}_{k}\psi(\bm x^{a}_i)$, and $\bm \phi^{a}_{k}$ are set to $\bm\phi^1=(2.0, 2.6)^\top$, $\bm\phi^2=(-1.0, -3.0)^\top$. The simplification of the generative mechanism is evident because of the linearity in the relationship between the covariates and the probability of the outcome, as well as the specification of parameters to enforce a clear separation among treatments and categories. 

The list of scenarios is designed as in the main simulation study. 

\begin{table}
\begin{tabular*}{\columnwidth}{@{}l@{\extracolsep{\fill}}|c@{\extracolsep{\fill}}c@{\extracolsep{\fill}}c@{\extracolsep{\fill}}|c@{\extracolsep{\fill}}c@{\extracolsep{\fill}}c@{\extracolsep{\fill}}c@{}}
  \hline\hline
 & \multicolumn{3}{c}{\textbf{Scenario S1a}} & \multicolumn{3}{c}{\textbf{Scenario S1b}} \\
& $MOT$ & $\%\Delta MTU$ & $NPC$ & $MOT$ & $\%\Delta MTU$ & $NPC$\\
  \hline
\multirow{2}{*}{pam-bp} & 5.8200 & 0.5995 & 14.7400 & 6.0200 & 0.5439 & 13.9400 \\ 
    & (2.7899) & (0.1691) & (2.9124) & (3.9096) & (0.2626) & (2.3248) \\[.05cm]
\multirow{2}{*}{km-bp}  & \bf{4.3200} & 0.6915 & \bf{14.9800} & \bf{5.0000} & 0.6490 & 12.9000 \\ 
    & (2.6298) & (0.1780) & (2.6991) & (2.7255) & (0.1880) & (2.4518) \\[.05cm]
\multirow{2}{*}{hc-bp}  & 6.6800 & 0.5671 & 14.5400 & 6.5000 & 0.5757 & 12.4800 \\ 
    & (1.7076) & (0.1369) & (2.7345) & (2.0429) & (0.1797) & (1.8653) \\[.05cm]
\multirow{2}{*}{dm-int} & 4.5400 & 0.7258 & 13.4800 & 5.0800 & \bf{0.7027} & 13.4600 \\ 
    & (3.5869) & (0.2219) & (2.8084) & (2.6484) & (0.1635) & (2.7346) \\[.05cm]
\multirow{2}{*}{t-ppmx} & 4.4000 & \bf{0.7491} & 14.8200 & 5.5000 & 0.6776 & \bf{14.4200} \\ 
    & (2.5153) & (0.1517) & (2.4302) & (3.0789) & (0.2122) & (2.7781) \\
   \hline
& \multicolumn{3}{c}{\textbf{Scenario S2a}} & \multicolumn{3}{c}{\textbf{Scenario S2b}} \\
& $MOT$ & $\%\Delta MTU$ & $NPC$ & $MOT$ & $\%\Delta MTU$ & $NPC$\\
  \hline
\multirow{2}{*}{pam-bp} & 5.5600 & 0.6112 & 9.4400 & 5.6800 & 0.5693 & 13.8400 \\ 
    & (2.9358) & (0.1847) & (2.3488) & (3.0199) & (0.2012) & (2.2710) \\ [.05cm]
\multirow{2}{*}{km-bp} & \bf{4.2600} & 0.7017 & 10.7800 & 5.0400 & 0.6417 & 13.0200 \\ 
 & (2.8198) & (0.1823) & (2.1120) & (2.3729) & (0.1634) & (2.3166) \\ [.05cm]
\multirow{2}{*}{hc-bp} & 7.0600 & 0.5302 & 11.6000 & 6.7600 & 0.5407 & 12.5200 \\ 
 & (2.1420) & (0.1796) & (2.6726) & (1.7907) & (0.1638) & (1.9713) \\ [.05cm]
 \multirow{2}{*}{dm-int} & 4.3600 & \bf{0.7376} & 13.4000 & \bf{4.9800} & 0.6962 & 13.7400 \\ 
    & (2.2656) & (0.1354) & (2.7180) & (3.0937) & (0.1994) & (2.6171) \\[.05cm]
\multirow{2}{*}{t-ppmx} & 5.0600 & 0.7111 & \bf{14.9800} & 5.1600 & \bf{0.6963} & \bf{14.9600} \\ 
 & (2.5667) & (0.1593) & (2.6146) & (3.0796) & (0.2129) & (2.5869) \\ 
   \hline
& \multicolumn{3}{c}{\textbf{Scenario S3a}} & \multicolumn{3}{c}{\textbf{Scenario S3b}} \\
& $MOT$ & $\%\Delta MTU$ & $NPC$ & $MOT$ & $\%\Delta MTU$ & $NPC$\\
  \hline
\multirow{2}{*}{pam-bp} & 5.8600 & 0.5983 & 14.2000 & 5.7400 & 0.5597 & 12.6400 \\ 
   & (3.6868) & (0.2398) & (2.2406) & (2.9194) & (0.2059) & (2.2475) \\ [.05cm]
\multirow{2}{*}{km-bp}  & 4.9400 & 0.6500 & 14.6000 & 4.9000 & 0.6512 & 12.9200 \\ 
    & (3.6668) & (0.2461) & (2.2314) & (3.3028) & (0.304) & (2.2932) \\[.05cm] 
\multirow{2}{*}{hc-bp}   & 7.3400 & 0.5167 & 14.1000 & 7.0000 & 0.5289 & 12.6800 \\ 
   & (1.7798) & (0.1473) & (2.5414) & (1.8844) & (0.1761) & (2.0247) \\ [.05cm]
\multirow{2}{*}{dm-int} & \bf{3.5600} & 0.7752 & 13.9200 & \bf{4.5400} & \bf{0.7263} & 13.7200 \\ 
& (2.2782) & (0.1381) & (3.1418) & (2.1685) & (0.1407) & (2.4993) \\[.05cm]
\multirow{2}{*}{t-ppmx}  & 3.6000 & \bf{0.7913} & \bf{15.0600} & 5.2600 & 0.6956 & \bf{14.7400} \\ 
 & (2.7180) & (0.1603) & (2.7435) & (2.5380) & (0.1521) & (2.3974) \\
   \hline
\end{tabular*}
\caption{Predictive performance: mean across $50$ replicated datasets, standard deviations are in parentheses. In each scenario and for each index the best performance is in bold.}
\label{tab:scenlin}
\end{table}

Table \ref{tab:scenlin} reports the results. The simpler generating mechanism results in overall better performances for all the methods, in particular we observe lower $MOT$ and larger $\%\Delta MTU$. 
Unsurprisingly, dm-int demonstrates good performances, since it is the only model that assumes a (log-) linear relationship between covariates and outcome probabilities. Nonetheless, t-ppmx confirms to be a flexible and robust model, regardless of the generating mechanism. Among the two-stage methods, km-bp is the one that yields the better results. 

T-ppmx outperforms all competing models in terms of $NPC$. With respect to $MOT$ and $\%\Delta MTU$, dm-int and t-ppmx present really close performances. Overall, t-ppmx attains better results when the number of covariates is moderate, while its predictive ability deteriorates as the number of predictive markers increase. Given that the two methods yield similar results, the great advantage provided by t-ppmx is that it allows for interpretable inference on the patients' clustering structure and identify patients more/less likely to benefit from a given treatment. 

\label{sec:ass}
\subsection{Recovering true clustering structure}\label{ssec:kcs}
We perform an additional comparative study to evaluate our method's ability to recover a latent structure in the data. We consider our predictive model with two different distributions for the random partition. In particular, we compare the proposed PPMx with a PPM employing the eppf induced by a DP as cohesion function. 
Note that such a PPM can be considered as a special case of our model, taking $\sigma = 0$ and $g(\bm x)\equiv 1$. Namely, we want to ascertain that the covariate-dependent prior we devised for the random partition adequately detects latent structure in the data.

We follow the generative mechanism described in Section 7.1, but instead of the data available from \cite{golub1999molecular}, we use synthetic data --whose latent structure is known-- as predictive biomarkers.

The first Scenario S1 closely follows \cite{argiento2022clustering}. In particular, we generate 4 covariates $(x_{i1}, \dots, x_{i4})$, for $i=1, \dots, n$. The first two covariates are continuous, while the last two are binary. Covariates are independently generated from three groups, with sizes 75, 75, and 50, respectively, as follows: $$(x_{i1}, x_{i2})\iid N_2(\bm\mu, 0.5I_2)~~x_{i3}, x_{i4}\iid Bern(q),$$ where $I_2$ is a $2\times 2$ identity matrix. 
For the first group $\bm \mu=(-3, 3)^\top$ and $q=0.1$. For the second one $\bm \mu=(0, 0)^\top$ and $q=0.5$, while for the third one $\bm \mu=(3, 3)^\top, q= 0.9$. 

Scenarios S2 and S3 follow the data generative mechanism 4 in \cite{page2018calibrating}. In particular, we consider $Q$ covariates independently generated from the following distributions: $20\%$ come from $N(0, 1)$, $20\%$ from $U(0, 10)$, $20\%$ from $t_4$, ($t$ distribution with $4$ degrees of freedom), $20\%$ more from $SN(10, 1, 10)$, (a skew-normal distribution) and the last $20\%$ from a two-component mixture of the form $0.4N(0, 1) +
0.6N(10, 2)$. Scenarios S2 and S3 differ for the number of predictive covariates, that is, $Q=10$ and $Q=20$, respectively. 

We use metrics presented in Section 7.2 to assess model prediction. We compute log pseudo-marginal likelihood (lpml) and Watanabe-Akaike Information Criterion (WAIC) to evaluate model fit. We generate $50$ datasets for each scenario. Each generated dataset has $n=200$ observations, with $170$ classified as training observations. 

\begin{table}[ht]
\centering
\begin{tabular}{rcccccc}
  \hline
 \multicolumn{7}{c}{\textbf{Scenario S1}} \\[.1cm]
 & $MOT$ & $\%\Delta MTU$ & $NPC$ & lpml & $ARI^1$ & $ARI^2$ \\[.1cm]
 \hline
 \hline
\multirow{2}{*}{PPMx} &	9.9600 & 0.4107 & 16.6000 & -102.4659 & 0.9491 & 0.9608 \\
 &	(4.3657) &	(0.2893) &	(2.6030) &	(7.2650) &	(0.0380) &	(0.0578)\\[.15cm]
\multirow{2}{*}{PPM} &	10.6800 &	0.3159 & 15.5600 &	-156.5070 &	- & -\\
 &	(4.551) &	(0.3123) &	(2.4173) &	(49.4241) &	- &	-\\
  \hline
  \multicolumn{7}{c}{\textbf{Scenario S2}} \\[.1cm]
 & $MOT$ & $\%\Delta MTU$ & $NPC$ & lpml & $ARI^1$ & $ARI^2$ \\[.1cm]
 \hline
 \multirow{2}{*}{PPMx} &	6.9200 & 0.7387 & 17.0000 &	-101.4985 &	0.5206 & 0.4401 \\
 &	(4.1837) &	(0.3581) &	(2.5395) &	(30.3342) &	(5.1251) & (0.1452) \\[.15cm]
\multirow{2}{*}{PPM} &	11.0800 &	0.3941 &	15.7800 &	-143.8411  &	- &	- \\
 &	(6.7788) &	(0.5999) &	(2.8161) &	(39.9999) &	- &	- \\
 \hline
  \multicolumn{7}{c}{\textbf{Scenario S3}} \\[.1cm]
 & $MOT$ & $\%\Delta MTU$ & $NPC$ & lpml & $ARI^1$ & $ARI^2$ \\[.1cm]
 \hline
 \multirow{2}{*}{PPMx} &	10.0200 & 0.3742 & 15.1800 & -104.5730 &0.5825 & 0.5464 \\
 &	(4.7915) &	(0.3592) &	(2.8190) &	(6.5637) &	(0.0925) &	(0.1039) \\[.15cm]
\multirow{2}{*}{PPM} &	10.8600 &	0.2984 & 15.2000 &	-157.9971 &	- &- \\
 &	(6.4047) &	(0.4783) &	(2.6573) &	(45.9023) &	- &- \\
\hline
\end{tabular}
\caption{Prediction performances, model fit measures, and ARI index for Treatment 1 and 2 for Scenarios S1-S3: mean across 50 replicated datasets (standard deviations are in parentheses).}
\label{tab:addsim}
\end{table}

We report averaged results in Table \ref{tab:addsim}, with standard deviations in parenthesis. To evaluate our method's ability to recover a latent structure in the data, we also report ARI indices for PPMx attained in the two treatments.
PPMx consistently outperforms PPM with respect to metrics evaluating prediction. PPM's poor performance is due to the response variable not being an explicit function of the covariates. Therefore, PPM can not recover any structure on the predictive covariates. For this reason, ARI is not computed for PPM. 

\subsection{Misspecification}\label{ssec:misspec}
A simulation study is conducted to assess the robustness of the method in the face of potential misspecification of prognostic and predictive covariates. To investigate the impact of missclassification on both clustering and predictive performance, two scenarios (S4 and S5) are constructed based on S2 and S3, where the clustering structure of the data is known. Specifically, building on Scenario S2, Scenario S4 was generated by designating two randomly selected predictive covariates as prognostic, and one of two randomly selected prognostic covariates as predictive. Similarly, Scenario S5 was created from Scenario S3 by randomly exchanging five of the 20 predictive covariates with one of the two prognostic covariates. 

\begin{table}[ht]
\centering
\begin{tabular}{rcccccc}
  \hline
 & $MOT$ & $\%\Delta MTU$ & $NPC$ & lpml & $ARI^1$ & $ARI^2$ \\[.1cm]
 \hline
 \hline
\multirow{2}{*}{\textbf{Scenario S4}} &	10.4600 & 0.4686 & 15.0400 & -110.8869  & 0.4015 & 0.4159 \\
 &	(5.4369) &	(0.4734) &	(2.6416) &	(6.9667) &	(0.1713) &	(0.1573)\\[.15cm]
 \hline
 \multirow{2}{*}{\textbf{Scenario S5}} &	11.7600 & 0.2571  & 12.9000 & -113.1016 & 0.5682 & 0.5705 \\
 &	(4.5828) &	(0.3621) &	(3.1249) &	(10.6491) &	(0.1849) &	(0.2023) \\[.15cm]
 \hline
\end{tabular}
\caption{Prediction performances, model fit measures, and ARI index for Treatment 1 and 2 for Scenarios S4-S5: mean across 50 replicated datasets (standard deviations are in parentheses).}
\label{tab:addsim2}
\end{table}

T-ppmx shows to be quite robust to the misspecification of prognostic/predictive covariates. Specifically, when comparing the outcomes presented in Table \ref{tab:addsim2} with those observed in Table \ref{tab:addsim} under scenarios S2 and S3, where no misspecification is present, it becomes evident that the t-ppmx approach is capable of recovering the clustering structure to a comparable extent. However, the results appear to be primarily impacted in terms of their predictive performance. In fact, the $\%\Delta MTU$ metric is considerably more affected by misspecification. Notably, the impact of misspecification is less severe in scenarios featuring a greater number of predictive covariates. This is exemplified by Scenario S5, which experiences a less significant loss with respect to Scenario S3, in contrast to the more pronounced effects seen comparing S4 to S2.

\section{Posterior Inference}
\label{sec:pi}
We describe the MCMC algorithm to simulate from the posterior distribution of the parameters of interest:
\begin{equation}
\begin{split}
p(\bm{\eta}^{\star}, \bm{\mathcal{P}},  \bm{\pi}, \bm{\beta}|\bm{y},\bm{x},\bm{z}) = 
\prod_{a=1}^T p(\bm{\eta}^{a\star}, {\mathcal{P}}^{a}_{n^a}, \bm{\pi}^a, \bm{\beta}|\bm{y}^a,\bm{x}^a,\bm{z}^a), \\
p(\bm{\eta}^{a\star}, \mathcal{P}^{a}_{n^a}, \bm{\pi}^a, \bm{\beta}|\bm{y}^a,\bm{x}^a,\bm{z}^a) =  
\prod_{i=1}^{n_a} {\pi^a_{i y^{a}_{i}}} \prod_{j=1}^{C^{a}_{n^a}} \prod_{i \in S_j^a}
\text{Dirichlet}(\bm\pi_{i}^{a}|\bm\gamma_i^a(\bm\eta_{j}^{a\star},\bm{\beta})).
\end{split}
\label{eq:Hpost}
\end{equation}
We adopt a data augmentation approach \citep{argiento2022infinity} to represent the Dirichlet distribution as independent latent Gamma random variables. In particular, we reparameterize equation 1 in the paper letting $\pi_{ik}^a=d_{ik}^a/D_{i}^{a}$, where $D_{i}^{a}=\sum_{k=1}^{K}d_{ik}^{a}$ and assume that $d_{ik}^{a}\sim Gamma(\gamma_{ik}^{a}(\eta_{jk}^{a\star}, \bm\beta_k), 1)$. More details in Section \ref{ssec:ui}. 

The core part of the MCMC algorithm is the update of cluster membership. The computation associated with 
the joint law of $(\mathcal{P}_{n^a}^a, \bm\eta_{j}^{a\star})$ is based on \cite{neal2000markov}'s Algorithm 8 with a reuse strategy \citep{favaro2013mcmc}. Conditional on the updated cluster labels, all the remaining parameters are easily updated with Gibbs sampler or Metropolis-Hastings steps. In the following, we outline the implemented Metropolis-Hastings within Gibbs algorithm: 
\begin{itemize}
    \item[1.]\textbf{update} $\{\mathcal{P}^{a}_{n^a}\}$ by \cite{neal2000markov}'s Algorithm 8 coupled with a reuse strategy \cite{favaro2013mcmc};
    \item[2.] \textbf{update} $\{\bm{\eta}_{j}^{a\star}\}$ by Metropolis-Hastings. Hyperparameters are updated through Gibbs steps from their respective full conditional distributions;
    \item[3.] \textbf{update} $\{(\kappa^{a}, \sigma^{a})\}$ evaluating $p((\kappa^a, \sigma^a)|\mathcal{P}_{n^a}^{a}, \cdot)$ at a finite, discrete grid of possible $(\kappa^a, \sigma^a)$ values; \item[4.] \textbf{update} $\{\beta_{pk}\}$ by Metropolis-Hastings. Hyperparameters are updated through a slice sampler \citep{neal2003slice} as suggested in \cite{polson2014bayesian};
    \item[5.] \textbf{update} $\{d^{a}_{ik}\}$ drawing from its full conditional distributions, i.e. using Gibbs sampler.
\end{itemize}
A full discussion on the Metropolis within Gibbs we implemented is given in Section \ref{ssec:steps}.
Finally in Section \ref{sec:postclu} we discuss our approach for posterior clustering.
\subsection{Augmented data scheme}\label{ssec:ui}
Generating samples from the Dirichlet distribution using independent Gamma random variables is computationally efficient. Exploiting this property, we adopt a data augmentation approach based on a reparameterization of equation 1 in the paper and the introduction of an auxiliary parameter. We first assume that the response of the $i-$th patient to the treatment $a$ follows a Multinomial distribution:
$$y_{i}^{a}|\bm{\pi}_{i}^a \ind \text{Multinomial}(1,\bm{\pi}_{i}^{a}).$$ We assume a conjugate prior on the response probability 
$$\bm \pi_{1}^a,\dots,\bm \pi_{n^a}^a\mid\bm\eta_1^{a\star},\dots,\bm{\eta}_{C^{a}_{n^a}}^{a\star},\mathcal{P}_{n^a}^{a},\bm \beta \sim  \prod_{j=1}^{C^{a}_{n^a}}\prod_{i\in S^{a}_{j}}\text{Dirichlet}(\bm \gamma_i^a(\bm \eta_{j}^{a\star},\bm \beta)).$$

We introduce latent random variables $d_{ik}^a\sim\text{Gamma}(\gamma_{ik}^a(\eta_{jk}^{a\star}, \bm\beta_k), 1)$ constructed such that $\pi_{ik}^{a}=d_{ik}^{a}/D_{i}^{a}$, where $D_{i}^{a}=\sum_{k=1}^{K}d_{ik}^{a}$, obtaining
$$y_{i}^{a}|\bm d_{i}^a, D_{i}^a \ind \text{Multinomial}(1,\bm{d}_{i}^{a}/D_{i}^{a}),$$
where $\bm{d}_{i}^{a}=(d_{i1}^{a}, \dots, d_{iK}^{a})^\top$.

Quantity $p(\bm{\eta}^{a\star}, {\mathcal{P}}^{a}_{n^a}, \bm{\pi}^a, \bm{\beta}|\bm{y}^a,\bm{x}^a,\bm{z}^a)$ in equation \eqref{eq:Hpost} can be restated as
\begin{equation}\label{eq:Hpost2}
p(\bm{\eta}^{a\star}, {\mathcal{P}}^a_{n^a}, \bm{d}^a, \bm{\beta}\mid\bm{y}^a,\bm{x}^a, \bm{z}^a)=\prod_{i=1}^{n_a} \frac{d_{iy^{a}_{i}}^a}{D_i^a} \prod_{j=1}^{C^a_{n^a}} \prod_{i \in S_j^a}
\text{Gamma}(\bm\gamma_i^a(\bm\eta_{j}^{a\star}, \bm\beta_k),1),
\end{equation}
For $a=1, \dots, T$, we introduce $n^a$ auxiliary parameters $u_{i}^{a}$ and let $u_i^a\mid D_i^a\sim\text{Gamma}(1, D_i^a)$.
From the Gamma density function we obtain that
\begin{equation*}
\frac{1}{D_i^a}=\int_{0}^{\infty}\exp(-D_i^a u_i^a)\wrt u_i^a,
\end{equation*}
so from equation \eqref{eq:Hpost2}:
\begin{flalign*}
p(\bm{\eta}^{a\star}, {\mathcal{P}}^a_{n^a}, \bm{d}^a, \bm{\beta}, \bm u\mid\bm{y}^a,\bm{x}^a, \bm{z}^a)
\nonumber &=\prod_{i=1}^{n^a} d_{iy^{a}_{i}}^a \mathrm{e}^{-D_i^au_i^a} \Bigg(\prod_{j=1}^{C_{n^a}^a} \prod_{i \in S_j^a}\prod_{k=1}^K\dfrac{{d_{ik}^a}^{\gamma_{ik}^a(\eta_{jk}^{a\star}, \bm\beta_k)-1}\mathrm{e}^{-d_{ik}^a}}{\Gamma(\gamma_{ik}^a(\eta_{jk}^{a\star}, \bm\beta_k))}\Bigg)\\
\nonumber &=\prod_{i=1}^{n^a} d_{iy^{a}_{i}}^a \mathrm{e}^{-u_i^a\sum_kd_{ik}^a} \Bigg(\prod_{j=1}^{C_{n^a}^a} \prod_{i \in S_j^a}\prod_{k=1}^K\dfrac{{d_{ik}^a}^{\gamma_{ik}^a(\eta_{jk}^{a\star}, \bm\beta_k)-1}\mathrm{e}^{-d_{ik}^a}}{\Gamma(\gamma_{ik}^a(\eta_{jk}^{a\star}, \bm\beta_k))}\Bigg)\\
\nonumber &=\prod_{i=1}^{n^a} d_{iy^{a}_{i}}^a \prod_{k=1}^K\mathrm{e}^{-u_i^a d_{ik}^a} \Bigg(\prod_{j=1}^{C_{n^a}^a} \prod_{i \in S_j^a}\prod_{k=1}^K\dfrac{{d_{ik}^a}^{\gamma_{ik}^a(\eta_{jk}^{a\star}, \bm\beta_k)-1}\mathrm{e}^{-d_{ik}^a}}{\Gamma(\gamma_{ik}^a(\eta_{jk}^{a\star}, \bm\beta_k))}\Bigg)\\
\nonumber &= \prod_{i=1}^{n^a} d_{iy^{a}_{i}}^a \Bigg(\prod_{j=1}^{C_{n^a}^a} \prod_{i \in S_j^a}\prod_{k=1}^{K}\dfrac{{d_{ik}^a}^{\gamma_{ik}^a(\eta_{jk}^{a\star}, \bm\beta_k)-1}\mathrm{e}^{-d_{ik}^a(u_i^a+1)}}{\Gamma(\gamma_{ik}^a(\eta_{jk}^{a\star}, \bm\beta_k))}\Bigg).
\end{flalign*}
\begin{center}
\end{center}
\subsection{MCMC sampling}
\label{ssec:steps}
For the posterior inference, we designed a Metropolis within Gibbs sampler. We use Algorithm 8 by \cite{neal2000markov} with a reuse strategy \citep{favaro2013mcmc} for cluster label updates. 

\noindent
We introduce here the latent cluster allocations. In particular, for $a=1, \dots, T$, let $\bm{e}^a = (e_1^a, \ldots, e_{n^a}^a)^\top$ be the cluster allocation vector of indexes, with $e_i^a = j$ iff $i \in S_j^a$. Conditional on the updated cluster labels, all the remaining parameters are updated with Gibbs sampler or Metropolis-Hastings steps. We briefly outline the scheme in Algorithm \ref{alg:mcmc}, and we give full details in the following. 

\begin{algorithm}[h]
\caption{MCMC sampling}
\label{alg:mcmc}
\DontPrintSemicolon
\For{$l = 1, \dots, niter$} 
{
  \For{$a = 1, \dots, T$}    
        {
         \tcc{Algorithm 8 with Reuse}
         Update the random partition $\mathcal{P}^{a}_{n^a}$\;
         \For{$a = j, \dots, C_{n^a}^a$}    
            {
                \tcc{Metropolis step}
                Update the cluster specific parameters $\bm\eta_{j}^{a\star}$\;
            }
        \tcc{Gibbs step}
        Update the hyperparameters of the cluster specific parameters $(\bm\theta^a, \bm\Lambda^a)$\;
        \tcc{Gibbs step}
        Update the NGGP parameters $(\kappa^a, \sigma^a)$ \;
    }
    \tcc{Metropolis step}
    Update prognostic coefficients $\{\beta_{pk}\}_{p=1, \dots, P, k=1, \dots, K}$\;
    \tcc{Slice sampler}
    Update hyperparameters of the prognostic coefficients $\{\lambda_{pk}\}_{p=1, \dots, P, k=1, \dots, K}$ and $\{\tau_{k}\}_{k=1, \dots, K}$\; 
    \For{$a = 1, \dots, T$}    
        {
            \tcc{Gibbs step}
            Update latent random variable $\{d_{ik}^a\}_{i=1, \dots, n^a, k=1, \dots, K}$\;
            \tcc{Gibbs step}
            Update auxiliary parameters$\{u_{i}^a\}_{i=1, \dots, n^a}$\;
        }
}
\end{algorithm}

\begin{itemize}
\item[$\mathcal{P}^{a}_{n^a}$:] \emph{Algorithm 8 with Reuse.} For $a=1,\ldots, T$, and $i=1,\ldots, n^a$, let $S_j^{a,-i}$ and $C^{a,-i}_{n^a}$ denote the $j$-th cluster and the total number of clusters when subject $i$ assigned to treatment $a$ is not considered. In the same way, we use $\bm x_j^{a \star, -i}$ to denote the matrix of predictive determinants of the patients in cluster $j$ when the $i-$th patient is not included. Cluster membership for patient $i$ that is $e_{i}^{a}$ is drawn using the following unnormalized probabilities:
\begin{multline}
\label{eq:prob}
P(e_i^a = j|\cdot) \propto \\
\begin{dcases*}
\prod_{k=1}^{K}\dfrac{{d_{ik}^a}^{\gamma_{ik}^a(\eta_{jk}^{a\star}, \bm\beta_k)-1}\mathrm{e}^{-d_{ik}^a(u_i^a+1)}}{\Gamma(\gamma_{ik}^a(\eta_{jk}^{a\star}, \bm\beta_k))} \frac{\rho(S_{j}^{a, -i} \cup \{ i \})  g(\bm{x}_j^{a \star, -i} \cup \{\bm{x}^a_i\})}{\rho(S_{j}^{a, -i} ) g(\bm{x}_j^{a \star, -i})}~ \text{ for }  j = 1,\ldots, C^{a, -i}_{n^a}\\
\prod_{k=1}^{K}\dfrac{{d_{ik}^a}^{\gamma_{ik}^a(\eta_{jk}^{a\star}, \bm\beta_k)-1}\mathrm{e}^{-d_{ik}^a(u_i^a+1)}}{\Gamma(\gamma_{ik}^a(\eta_{jk}^{a\star}, \bm\beta_k))}\frac{\rho(\{ i \}) g(\{\bm{x}^a_i\})}{M}~ \text{ for } j = C^{a, -i}_{n^a} +1,\ldots, C^{a, -i}_{n^a} + M, \\
\end{dcases*}
\end{multline}

where $\{\eta_{j k}^{a\star}\}$ for $j=C^{a, -i}_{n^a} +1,\ldots, C^{a, -i}_{n^a} + M$ are $M$ auxiliary variables \citep{neal2000markov}, associated with $M$ empty clusters, independently and identically distributed according to some prior distribution $p^e$. 
The first terms are the likelihoods associated with observation $i$ given the clusters' parameters. The second terms can be interpreted as being proportional to the covariate-informed prior probability of being assigned to the corresponding cluster (with the $M$ empty clusters sharing the probability of creating a new cluster).

Algorithm 8 with Reuse proposes efficient handling of the $M$ auxiliary parameters; it updates the cluster assignment of observation $i$ according to the following scheme reported in Algorithm \ref{alg:reuse}.

\begin{algorithm}[h]
\caption{Algorithm 8 with Reuse}
\label{alg:reuse}
\DontPrintSemicolon
  \For{$a = 1, \dots, T$}    
        {
            \For{$i = 1, \dots, n^a$}    
        {
        Remove $i$ from the cluster it belongs, so that $S_{j}^{a}\in\mathcal{P}_{n}^{a}$, $|S_{j}^{a}| = n^{a}_j$ becomes $S_{j}^{a, -i}, |S_{j}^{a}| = n^{a}_j-1$ \;
            \If{$|S_{j}^{a}| = 0$\tcc{Cluster $S^{a}_{j}$ is empty}} 
    {
        Relabel $\bm S^{a, -i}$ and $\bm e^{a, -i}$ to avoid gaps in the cluster labels\;
        Sample $m\in\{1, \dots, M\}$ uniformly at random \;
        Replace $\bm \eta_{C^{a, -i}_{n^a} +m}^{a\star}$ with $\bm \eta_{j}^{a\star}$ \; 
        Remove $S^{a}_{j}$ from $\mathcal{P}^{a}_{n^a}$ \;
    }
    Assign $i$ to the clusters with probabilities $P(e_{i}^{a}=j\mid\cdot)$ \;
    \If{$e_{i}^{a}\in\{C_{n^a}^{a, -i}+1, \dots, C_{n^a}^{a, -i}+M\}$ \\
  \tcc{The observation $i$ is assigned to an empty cluster}
  }
    {
    	Assign it to a new cluster in $\mathcal{P}^{a}_{n}$ with parameter $\bm\eta^{a\star}_{e_{i}^{a}}$ \;
        Replace $\bm\eta^{a\star}_{e_{i}^{a}}$ with a new independent draw from $p^e$.\;
    }
    }
    \tcc{Now we update empty clusters' parameters with the Reuse strategy}
    Discard $\{\bm\eta_{j}^{a\star}\}$ for $j=C_{n^a}^{a}+1, \dots, C_{n^a}^{a}+M$ \;
    Sample $M$ iid $\{\bm\eta_{j}^{a\star}\}$s from $p^e$ (for $j=C_{n^a}^{a}+1, \dots, C_{n^a}^{a}+M$) 
    }
\end{algorithm}

In the original formulation of Algorithm 8 \citep{neal2000markov}, after the update of the cluster label, the auxiliary parameters associated with empty clusters are discarded. Since after the cluster update for the $i-$th subject, the parameters for unused empty clusters are already independently and identically distributed, \cite{favaro2013mcmc} propose to reuse them for the update of the subsequent observations. Note that the only difference adopting the Reuse Algorithm implies is the way the parameters of the empty clusters are managed and retained across cluster assignment updates of multiple observations.

Finally, particular attention should be paid to the relabeling step at line 6 of Algorithm \ref{alg:reuse}. In fact, to avoid gaps in the cluster labels we need to relabel all the clusters \citep{page2015predictions}. We denote with $\bm e^{a, -i}=(e_{1}^{a}, \dots, e_{n^a-1}^{a})^\top$ the cluster allocation vector of indices when subject $i$ assigned to treatment $a$ is not considered. We relabel $\{S_{j'}^{a, -i} : S_{j'}^{a, -i} > S_{j}^{a}\}$ for $j'=1, \dots, C^{a, -i}_{n^a}$. In particular, if $S_{j'}^{a, -i}>S_{j}^{a}$, $S_{j'}^{a, -i}=S_{j'}^{a, -i}-1$, for $j'=1, \dots, C^{a, -i}_{n^a}$.
Similarly, we relabel $\{e_{i'}^{a, -i} : e_{i'}^{a, -i} > e_{i}^{a}\}$ for $i'=1, \dots, n^a-1$. In particular, if $e_{i'}^{a, -i}>e_{i}^{a}$, $e_{i'}^{a, -i}=e_{i'}^{a, -i}-1$, for $i'=1, \dots, n^a-1$.

\item[$\bm{\eta}_{j}^{a\star}$:] \emph{Metropolis step.} For $a=1,\ldots, T$, $j=1,\ldots, C^{a}_{n^a}$ we propose ${\bm{\eta}^{a \star}_{j}}^{'}$ obtained from a random walk proposal and accept it with probability

\begin{equation*}
    \min\Bigg\{\dfrac{p(\bm d^a\mid{\bm{\eta}^{a \star}_{j}}^{'}, \cdot)p({\bm{\eta}^{a \star}_{j}}^{'})}{p(\bm d^a\mid\bm{\eta}^{a \star}_{j}, \cdot)p(\bm{\eta}^{a \star}_{j})}, 1\Bigg\}.
\end{equation*}

\textbf{Hyperparameters update}:
\begin{itemize}
    \item[$\bm \theta^{a}, \bm\Lambda^{a}$] \emph{Gibbs step.} For $a=1, \dots, T$ we sample $\bm \theta^{a}$ and $\bm\Lambda^{a}$ from their respective full conditionals:
    \begin{equation*}
        \bm \theta^{a}| \bm \Lambda^{a}, \bm\eta_{1}^{a\star}, \dots, \bm \eta_{j}^{a\star} \sim N_K\bigg(\dfrac{\bm \mu_{0}\nu_0 + n^{a}}{n^{a} + \nu_0}, \big(\bm \Lambda^{a}(n^{a}+\nu_0)\big)^{-1}\bigg),
    \end{equation*}
    \begin{equation*}
        \bm \Lambda^{a}| \bm\eta_{1}^{a\star}, \dots, \bm \eta_{j}^{a\star} \sim W\bigg(s_0+\frac{n^{a}}{2}, \bm\Lambda_0+\frac{n^{a}}{2}\big(\bar{\bm \Lambda}^{a} + \frac{s_0}{s_0+n^{a}}(\bar{\bm\eta}^{a}-\bm\mu_0)(\bar{\bm\eta}^{a}-\bm\mu_0)^\top\big)\bigg),
    \end{equation*}
    where $\bar{\bm{\eta}} = (\bar{\eta}_{1}, \dots, \bar{\eta}_{K})^\top$, $\bar{\eta}_{k}=\frac{1}{n}\sum_{j=1}^{C^{a}_{n^a}}\eta^{a\star}_{jk}n_j$, and $\bar{\bm\Lambda}=\frac{1}{n}\sum_{i=1}^{n}(\bm\eta_i-\bar{\bm\eta})(\bm\eta_i-\bar{\bm\eta})^\top$.
    
\end{itemize}

\item[$(\kappa^a, \sigma^a)$:] \emph{Gibbs step.} Since the normalizing constant of equation (4.3) can not be computed, we produce samples from a discretized approximation of posterior distribution, evaluating $p((\kappa^a, \sigma^a)|\mathcal{P}_n^a, \cdot)$ at a finite, discrete grid of possible $(\kappa^a, \sigma^a)$ values. In particular, we construct a $10\times 10$ grid in $(0, 15)\times (0.0, 0.6)$,such that the marginal
distributions are $\kappa\sim Gamma(2, 1)$ and $\sigma\sim Beta(5, 23)$, respectively. For $a = 1, \dots, T$, we evaluate (4.3) at each grid point, obtaining a discrete approximation of the log posterior distribution, and then we normalize the values, obtaining weights that sum to $1$ across the grid's points. Finally, we sample a new value for $(\kappa^a, \sigma^a)$ from the grid with respect to their corresponding normalized posterior probability.

\item[$\beta_{pk}$:] \emph{Metropolis step.} 
For $k=1,\ldots, K$,  we propose $\beta_{pk}$ from a random walk proposal and accept it with probability:

\begin{equation*}
    \min\Bigg\{\dfrac{p(\bm d^a\mid\beta^{'}_{pk}, \cdot)p(\beta^{'}_{pk})}{p(\bm d^a\mid\beta_{pk}, \cdot)p(\beta_{pk})}, 1\Bigg\}.
\end{equation*}




\textbf{Hyperparameters update}:
\begin{itemize}
    \item[$\lambda_{pk}$] The global scale parameters are updated through an adaptation of the slice sampling scheme given in the online supplement of \cite{polson2014bayesian}. For $p=1, \dots, P$ and $k=1, \dots, K$, we define $\varpi_{pk} = 1/\lambda_{pk}^2$ and $\varsigma_{pk}=\beta_{pk}/\tau_{k}$.
    This reparameterization allows us to employ slice sampler \citep{neal2003slice}, as the conditional posterior distribution of $\varpi_{pk}$ is
    \begin{equation*}
      p(\varpi_{pk}\mid \tau_{k}, \varsigma_{pk}) \propto \exp\Bigg\lbrace-\frac{\varsigma_{pk}^{2}}{2}\varpi_{pk}\Bigg\rbrace\frac{1}{1+\varpi_{pk}}.
    \end{equation*}
    To sample $\lambda_{pk}$:
    \begin{itemize}
      \item[1.] draw a sample from Uniform distribution:
      \begin{equation*}
        h_{pk}\mid \varpi_{pk} \sim U(0, 1/(1+\varpi_{pk}));
      \end{equation*}
      \item[2.] draw a sample from Truncated Exponential density, so that it has zero probability outside the interval $(0, (1-u_{pk})/u_{pk})$:
      \begin{equation*}
        \varpi_{pk}\mid \varsigma_{pk}, h_{pk} \sim~ Exp(2/\varsigma_{pk}^{2}).
      \end{equation*}
    \end{itemize}
    Transforming back to the $\lambda-$scale it will ensure a sample from the conditional distribution of interest.
    
    \item[$\tau_k$] The same applies for $\tau_{k}$, replacing $\varpi = 1/\tau^{2}_{k}$ and $\varsigma_{k}^2=\sum_{p=1}^{P} \beta_{pk}^2/2$, for $k=1, \dots, K$.
\end{itemize}
\item[$d^{a}_{ik}$:] \emph{Gibbs step.} For $a=1,\ldots, T$, $i=1,\ldots, n^a$ and $k=1, \dots, K$ we sample $d_{ik}^{a}$ from:
\begin{equation*}
d_{ik}^{a} \mid\cdot\sim\text{Gamma}(\gamma_{ik}^{a}(\eta_{jk}^{a \star}, \bm{\beta}_k) + \bm{\delta}_1(y_i^a), (u_{i}^{a} + 1)^{-1}).
\end{equation*}
where $\bm{\delta}_1$ is a $K \times 1$ vector of zeros, with $y_{i}^{a}-$th element equal to 1. 

\item[$u_{i}^{a}$:] \emph{Gibbs step.} For $a=1,\ldots, T$, $i=1,\ldots, n^a$ we sample $u_i^a$ from:
\begin{equation*}
u_i^a \mid\cdot\sim\text{Gamma}(1, {D^{a}_{i}}^{-1}).
\end{equation*}
\end{itemize}
%
%
%
\subsection{Posterior Clustering}
\label{sec:postclu}
The main inferential goal of our method is treatment prediction. The covariate-dependent random partition model is employed to obtain homogeneous groups of patients. Posterior inference for the random partition $\mathcal{P}^{a}_{n^a}$ is affected by label switching. Since symmetric priors are chosen for both the random partition $\mathcal{P}^{a}_{n^a}$ and cluster-specific parameters $p^{\ast}(\bm\zeta^{a\star})$, their posterior distribution is invariant under permutations of the component indices. As a consequence, cluster labels are not identifiable. Nonetheless, prediction is not affected by label-switching because the posterior predictive distribution marginalizes over all possible partitions \citep{muller2011product}. 
Let $\tilde{\bm x}$ and $\tilde{\bm z}$ denote predictive and prognostic markers of a new patient. Following \cite{muller2011product}, we predict the response of the new untreated patient through an importance sampling strategy as follows: at each step of the MCMC algorithm, given the current partition we first assign a new patient to a cluster using the following probability weights 
\begin{equation}
\label{eq:probpostpred}
w^{a}_j\propto
\begin{dcases*}
\frac{\rho(S_{j}^{a} \cup \{ n^a + 1 \}) g(\bm{x}_j^{a \star} \cup \{\bm{\tilde x}^a\})}{\rho(S_{j}^{a} )  g(\bm{x}_j^{a \star})}~ \text{ for }  j = 1,\ldots, C^{a}_{n^a}\\
\rho(\{ n^a + 1 \}) g(\{\bm{\tilde x}^a\})~ \text{ for } j = C^{a}_{n^a} +1,\dots, C^{a}_{n^a} +M\\
\end{dcases*}
\end{equation}
and then we generate the cluster-specific parameters $\tilde{\bm\eta}_{j}$ weighted by $w_j$'s. Given these parameter values, we can predict the response probabilities of the new patient simply by applying the multinomial response model; see also \cite{page2016spatial} for a similar approach.  

\begin{algorithm}[h]
\caption{Posterior Predictive sampling}
\label{alg:pps}
\DontPrintSemicolon
  \For{$a = 1, \dots, T$}  
        {
        Compute $g(\tilde{\bm x}^{a\star})$ \;  
         \For{$j = 1, \dots, C_{n^a}$}    
        {
        Compute $g(\bm x^{a\star}_{j})$ \;
        }
        \For{$j = C_{n^a}+1, \dots, C_{n^a}+M$}    
        {
        Compute $g(\bm x^{a\star}_{j}\cup\{\tilde{\bm{x}}^{a}\})$ \;
        }
        \For{$j = 1, \dots, C_{n^a}+M$}    
        {
        Compute $w_j$ from equation \eqref{eq:probpostpred} \;
        }
        Sample $e_{\tilde{i}}^{a}$ from $Multinomial(\bm w^{a})$ \;
        \If{$e_{\tilde{i}}^{a}\in\{C_{n^a}^{a}+1, \dots, C_{n^a}^{a}+M\}$ \\
  \tcc{The observation $\tilde{i}$ is assigned to an empty cluster}
  }
    {
    	Assign it to a new cluster in $\mathcal{P}^{a}_{n^a}$ with parameter $\bm\eta^{a\star}_{e_{\tilde{i}}^{a}}$ obtained froma a new independent draw from $G_0$ \;
    }
        
    Compute $\{\log(\gamma_{\tilde{i}k}^{a}(\bm\eta^{a\star}_{e_{\tilde{i}}^{a}}, \bm\beta_{k})\}_{k=1, \dots, K}$ from equation (2) in the main paper \;
    Draw $\bm \pi^{a}_{\tilde{i}}$ from equation (1) in the main paper \;
    \For{$k = 1, \dots, K$}    
        {
            Sample $\tilde{y}$ from $p(\tilde{y}^a=k\mid \bm y^a, \bm z^a, \bm x^a, \tilde{\bm z}, \tilde{\bm x})$) 
        }
    }
\end{algorithm}

Finally, the marginal posterior distribution for the number of clusters $p(C^{a}_{n^a}|y^{a}_{1}, \dots, $ $y^{a}_{n^a}, \bm x_{1}^{a}, \dots, \bm x_{Q}^{a})$ is available as a byproduct of the posterior distribution for the random partition. To obtain a point estimate for $p(\mathcal{P}^{a}_{n^a}|y^{a}_{1}, \dots, y^{a}_{n^a}, \bm x_{1}^{a}, \dots, \bm x_{Q}^{a})$ we follow \cite{wade2018bayesian} and adopt a decision-theoretic approach that uses the variation of information loss function on the space of clusterings to estimate the optimal partition and to characterize its uncertainty.

\section{Generating Mechanism}
We closely follow the strategy devised in \cite{ma2016bayesian, ma2019bayesian} to generate the simulated datasets, i.e., we do not employ our model as the generative mechanism. To emulate the complex correlation patterns of genomics data, we obtain prognostic and predictive covariates from a leukemia dataset. The data available from \cite{golub1999molecular} provide gene expression levels from $5000$ genes collected across $38$ patients, of which $11$ were diagnosed with acute myelogenous leukemia and the remaining with acute lymphoblastic leukemia. To obtain scenarios with larger sample size, \cite{ma2016bayesian, ma2019bayesian} devised a procedure to expand the dataset, yielding $n=152$ patients ($38\times 4$) with $Q=90$ predictive and $P=2$ prognostic biomarkers. This data-processing procedure is presented in detail in the Supplementary material of \cite{ma2019bayesian}. The patients are assigned to $T=2$ competing treatment, and $K = 3$ levels of the ordinal-valued response variable are considered. Since the observed treatment endpoints were unavailable, the treatment response is generated using two continuation-ratio logistic functions (see Supplementary Material for more details). Finally, the optimal treatment for each simulated patient is determined as the inner product between the ordinal response probability and the response level utility weight $\bm \omega$. In particular, we set $\bm\omega=(0, 40, 100)^\top$ to make the ordinal response reflect the clinical importance of each level \citep{ma2016bayesian}.
\subsection{Generating treatment response}
\label{sec:gtr}
The data available from \cite{golub1999molecular} do not provide the observed treatment endpoints, so the treatment response is generated using two continuation-ratio logistic functions.

The first one takes predictive biomarkers as arguments:
\begin{equation}
    \label{eq:crl1}
    r_{k}(\bm x_{i}^{a})=\log\Bigg(\frac{Pr(y^{a}_{i}=k|\bm x^{a}_{i})}{p(y^{a}_{i}<k|\bm x^{a}_{i})}\Bigg)=\alpha^{a}_{k}+\bm\phi^{a}_{k}\psi(\bm x^{a}_i)^3,~\text{for}~i=1, \dots, n^{a},
\end{equation}
where $\psi(\cdot)$ is a one-dimensional function of the first two principal components, used to summarize the information carried by predictive markers. 
Response-level probabilities for prognostic features are defined through the second continuation-ratio logistic function:
\begin{equation}
    \label{eq:crl2}
    r^{\ast}_{k}(\bm z_{i}^{a})=\log\Bigg(\frac{Pr(y^{a}_{i}=k|\bm z^{a}_{i})}{p(y^{a}_{i}<k|\bm z^{a}_{i})}\Bigg)=\iota_{k}+\bm\chi_{k}\bm z^{a}_i,~\text{for}~i=1, \dots, n^{a}.
\end{equation}
\noindent
The coefficients $\alpha^{a}_{k}, \bm\phi^{a}_{k}, \bm\iota^{a}_{k}$ and $\bm\chi_k$, for $k=1, \dots, K-1$, are set to value that could produce realistic response rates. In particular, $\bm\alpha^1=(-0.5, -1.0)^\top$, $\bm\alpha^2=(0.7, -1.0)^\top$, $\bm\phi^1=(1.5, 2.0)^\top$, $\bm\phi^2=(-0.5, -1.0)^\top$, $\iota = (1.0, -0.5)^\top$, $\bm\chi^1=(1.0, 0.5)^\top$, and $\bm\chi^2=(0.7, 1.0)^\top$. The probabilities for each level of the ordinal response variable were generated as the pointwise product of \eqref{eq:crl1} and \eqref{eq:crl2} for each patient. That is, the true ordinal response probability (ORP) for response $k$ 
\begin{equation*}
    \label{eq:tot}
    ORP^{a}_{ik} =\dfrac{\omega_k\Big(r_{k}(\bm x^{a}_{i})r^{\ast}_{k}(\bm z^{a}_{i})\Big)}{\sum_{k=1}^{K}\omega_k\Big(r_{k}(\bm x^{a}_{i})r^{\ast}_{k}(\bm z^{a}_{i})\Big)},~i=1, \dots, n^a. 
\end{equation*}

\noindent
See \cite{ma2019bayesian} for more details. 


\section{Additional details of performance metrics}
\label{sec:mtu}
\subsection{Construction of $\% \Delta MTU$ index}

The extent to which a treatment is beneficial for each patient is heterogeneous, and the improvement offered by a therapy varies from patient to patient. In order to account for this heterogeneity, performances should be evaluated considering the relative utility gain. The relative gain in Treatment Utility, $\%\Delta MTU$ \citep{ma2016bayesian} allows measuring the overall benefit ensured by a treatment selection rule in the case of $T=2$ competing treatments. Denoting with $MTU^a(i)$ the mean treatment utility of treatment $a$ for patient $i$, we can obtain the differential treatment utility as $\Delta MTU(i)=MTU^1(i)-MTU^2(i)$. Considering the true optimal treatment $A(i)$ and denoting with $t(i)$ the treatment recommended by selection rule, we can construct the indicator function $\delta_{t(i)}(A(i))$ that is defined as:
$$\delta_{t(i)}(A(i))=
\begin{cases}
1&~\text{if}~t(i)=A(i)\\
-1&~\text{if}~t(i)\neq A(i).
\end{cases}$$

The sum of the true gains achieved by the selection rule is $$\Delta MTU=\sum_{i=1}^n\delta_{t}(i)(A(i))|\Delta MTU(i)|.$$ The maximum possible gain in mean treatment utility varies in each simulation scenario. To make performance comparable also across scenarios, we consider the proportion of the maximum possible gain in total mean treatment utility attained by the selection rule, that is:
$$\%\Delta MTU=\Delta MTU/\Delta MTU_{opt},$$ where $\Delta MTU_{opt}$ is the maximum possible total $MTU$, achieved when all patients are assigned to their optimal treatment. Finally, $\%\Delta MTU$ is bounded above by $1$ when it always recommends the optimal treatment, and $\%\Delta MTU=-1$ when it fails to select the optimal therapy for all the patients.

\subsection{Empirical Summary Measure (ESM)}
TCGA data do not provide the true optimal treatment, and only the $NPC$ measure, among those discussed in Section 6.1, can be used. 
We employ an empirical summary measure \citep[ESM,][]{song2004evaluating} to evaluate the relative increase in the population response rate attributable to a treatment allocation method compared to random allocation. Let $Y$ be the binary outcome variables, taking $0$ for non-respondents or $1$ for respondent patients. We define the treatment contrast as $\Delta(\bm X, \bm Z) = Pr(Y=1|A=2, \bm X, \bm Z)-Pr(Y=1|A=1, \bm X, \bm Z)$, where $A=\{1,2\}$ denote the non-targeted and targeted treatment, respectively. Indicating with $Pr(Y=1|A_r)$ the probability of being a respondent under a randomized treatment assignment, we obtain the relative increase in the population response rate under a personalized treatment selection rule as: 
\[
\begin{split}
ESM= &
\{Pr(Y=1|A=2, \Delta(\bm X, \bm Z)>0)\times Pr(\Delta(\bm X, \bm Z)>0)+ \\ &
Pr(Y=1|A=1, \Delta(\bm X, \bm Z)<0)\times Pr(\Delta(\bm X, \bm Z)<0)\}-Pr(Y=1|A_r),
\end{split}
\] 

where $Pr(Y=1|A=2, \Delta(\bm X, \bm Z)>0)$ and $Pr(Y=1|A=1, \Delta(\bm X, \bm Z)<0)$ can be estimated as the response rates for the subset of patients assigned by the proposed method to the treatment actually received; $Pr(\Delta(\bm X, \bm Z)>0)$ and $Pr(\Delta(\bm X, \bm Z)<0)$ can be empirically estimated as the proportion of patients who are recommended for targeted and standard treatment, respectively. Let $n^1$ and $n^2$ be the number of patients that received treatment $A=1$ and $A=2$, respectively; the weights $Pr(\Delta(\bm X, \bm Z)>0)$ and $Pr(\Delta(\bm X, \bm Z)<0)$ can be estimated as $n^1/n$ and $n^2/n$, respectively. Finally, $Pr(Y=1|A_r)$ is the overall response rate under randomization, and can be estimated as the sample proportion of responders.

Note that we based this summary measure on only two response categories, responders (CR) and non-responders (PD + PS), whereas we used all three levels of the ordinal outcome in the data analysis and to implement personalized treatment selection. 

\section{Data pre-processing in \cite{ma2019bayesian}} 
In our study, we analyzed data from \cite{ma2019bayesian}. As the TCGA data used in the study were collected from observational studies, measures were taken to avoid potential bias by matching patients based on baseline covariates such as tumor grade, gender, age, and initial year of pathological diagnosis (IYPD). To achieve this, they used the \texttt{R} package \texttt{MatchIt} \citep{stuart2011matchit} with default settings, resulting in 79 pairs of patients with standardized mean differences of 0.000, -0.050, 0.051, and 0.162 for tumor grade, gender, age, and IYPD, respectively. Obtained matches were reasonably satisfactory, as all final standardized mean differences were below the suggested cut-off value of 0.25 \citep{imai2008misunderstandings}. 
Moreover, in \cite{ma2019bayesian}, to identify potential prognostic and predictive features among the 173 protein expressions measured in the LGG dataset, univariate logistic regression models were fitted with a protein, the treatment, and their interaction as covariates using the \texttt{MASS} package in \texttt{R} \citep{ripley2013package}. A protein was considered as predictive (prognostic) if the $P$-value obtained from Wald's test for the interaction (main) effect was less than 0.1. Using this criterion, 23 proteins were selected as potential predictive features and five as potential prognostic features. The complete list of predictive proteins is reported in Table \ref{tab:list}. 

\begin{table}[h]
\centering
\begin{tabular}{cccccc}
  \hline
 ACVRL1 & Src & HSP70 & HER2 & PAI-1 & Smad \\ 
 FOXO3a & PRAS40 & Cyclin & SF2 & Bad & Lck \\
 Caspase-7-cleavedD198 & Paxillin & MYH11 & $14-3-3-\epsilon$ & Akt & Caveolin-1 \\
 Rab25 & YAP & RBM15 & Claudin-7 & ER-$\alpha$ & C-Raf \\
 CD31 & Ku80 & Bcl-2 & GSK3-$\alpha$-$\beta$ &  &  \\
 
   \hline
\end{tabular}
\caption{List of predictive proteins used in the LGG case study.}
\label{tab:list}
\end{table}
          
          In the subsequent analysis, two prognostic covariates, namely \texttt{ACVRL1-R-C} and \texttt{HSP70-R-C}, were utilized as they exhibited the highest accuracy rate (78/158) in discriminant analysis. This analysis was conducted using 79 pairs of matched data with binary outcomes of PD/SD/PR as 0 and CR as 1. 

\section{MCMC Diagnostic Checks for LGG Data Analysis}
Convergence has been assessed through Gelman-Rubin potential scale reduction factor \citep[PSRF,][]{gelman1992inference} for parameters associated with prognostic covariates.

\begin{table}[h]
\centering
\begin{tabular}{rrrr}
  \hline
 & 1.0909 & 1.0173 & 1.1129 \\ 
 & 1.1159 & 1.1102 & 1.0991 \\ 
   \hline
\end{tabular}
\caption{Potential Scale Reduction Factor for prognostic covariates' regression coefficients $\{\beta_{pk}\}.$} 
\end{table}

\begin{figure}[h]
    \centering
    \includegraphics[width=.9\linewidth, keepaspectratio]{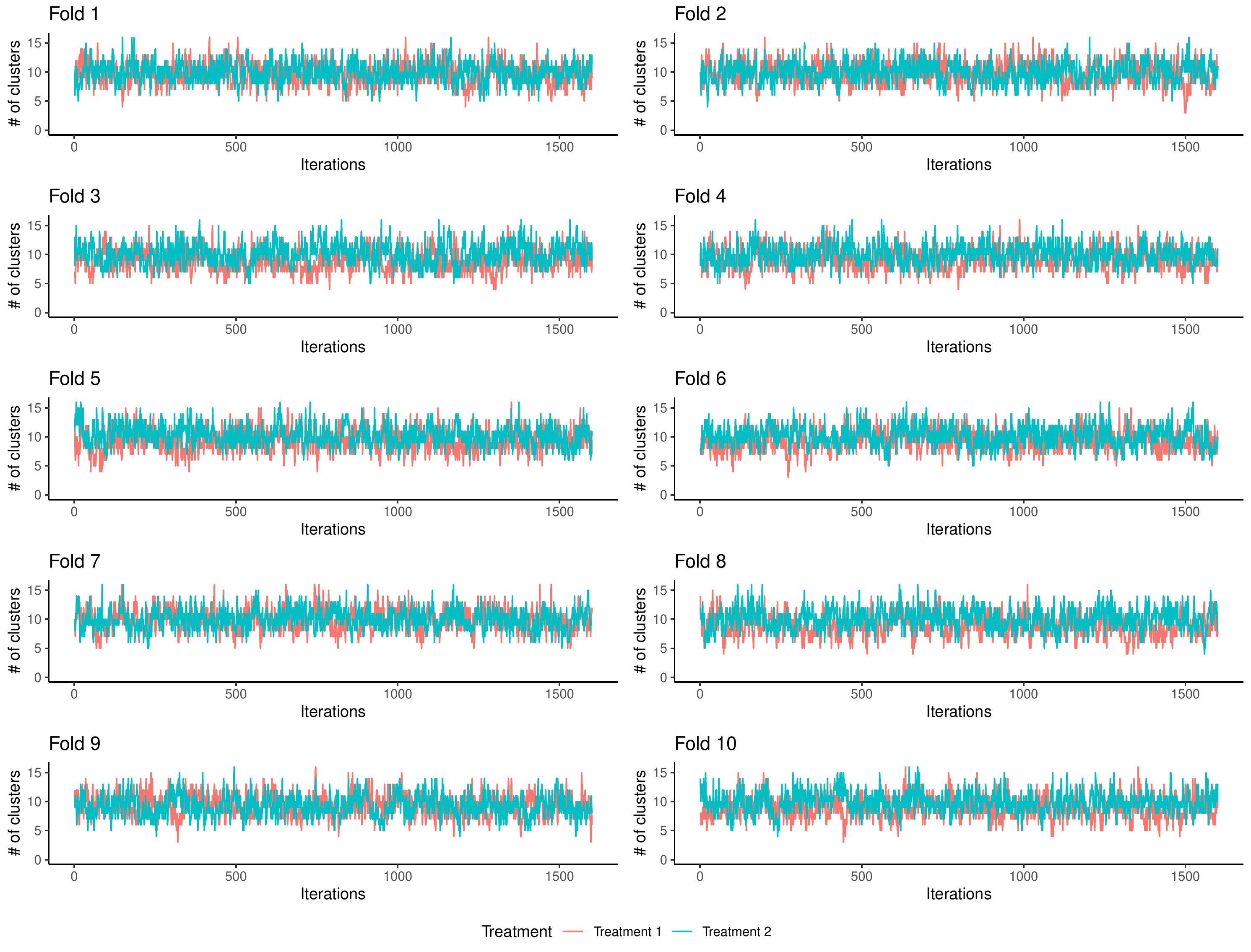}
    \caption{Traceplots of the number of clusters.}
    \label{fig:ncp}
\end{figure}

\begin{figure}[h]
    \centering
    \includegraphics[width=.9\linewidth, keepaspectratio]{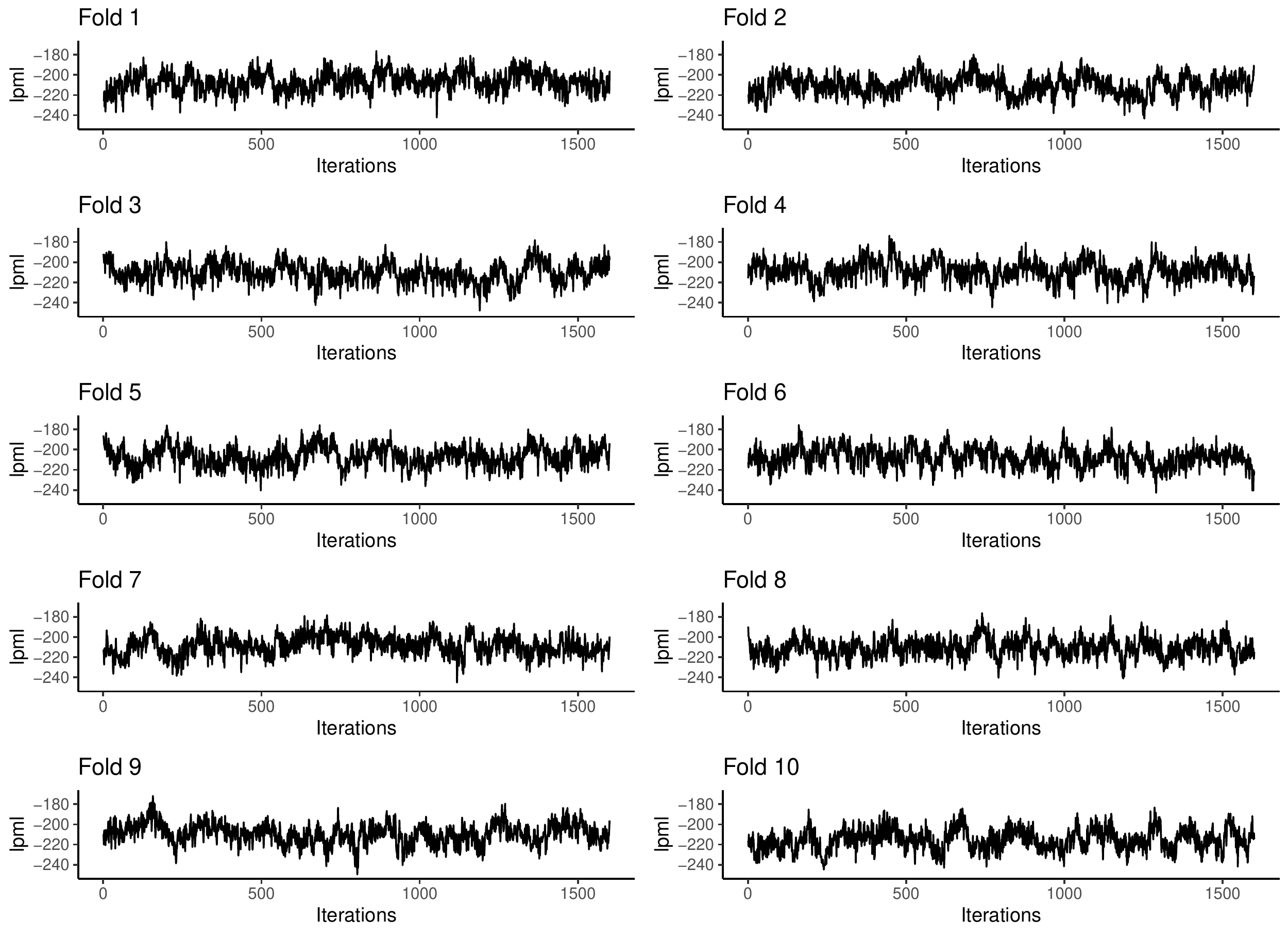}
    \caption{Traceplots of the lpml.}
    \label{fig:lpmlp}
\end{figure}

To assess the goodness-of-fit and the convergence of the parameters involved in the PPMx, we report in Figures \ref{fig:ncp} and \ref{fig:lpmlp} the trace plots of the number of clusters and the $lpml$, respectively. In both cases, the trace plots are displayed separately for each fold. 

These results do not raise any particular concerns on the MCMC convergence.

\section{Additional Figure}
\begin{figure}[h]
    \centering
    \includegraphics[width = .9\linewidth, keepaspectratio]{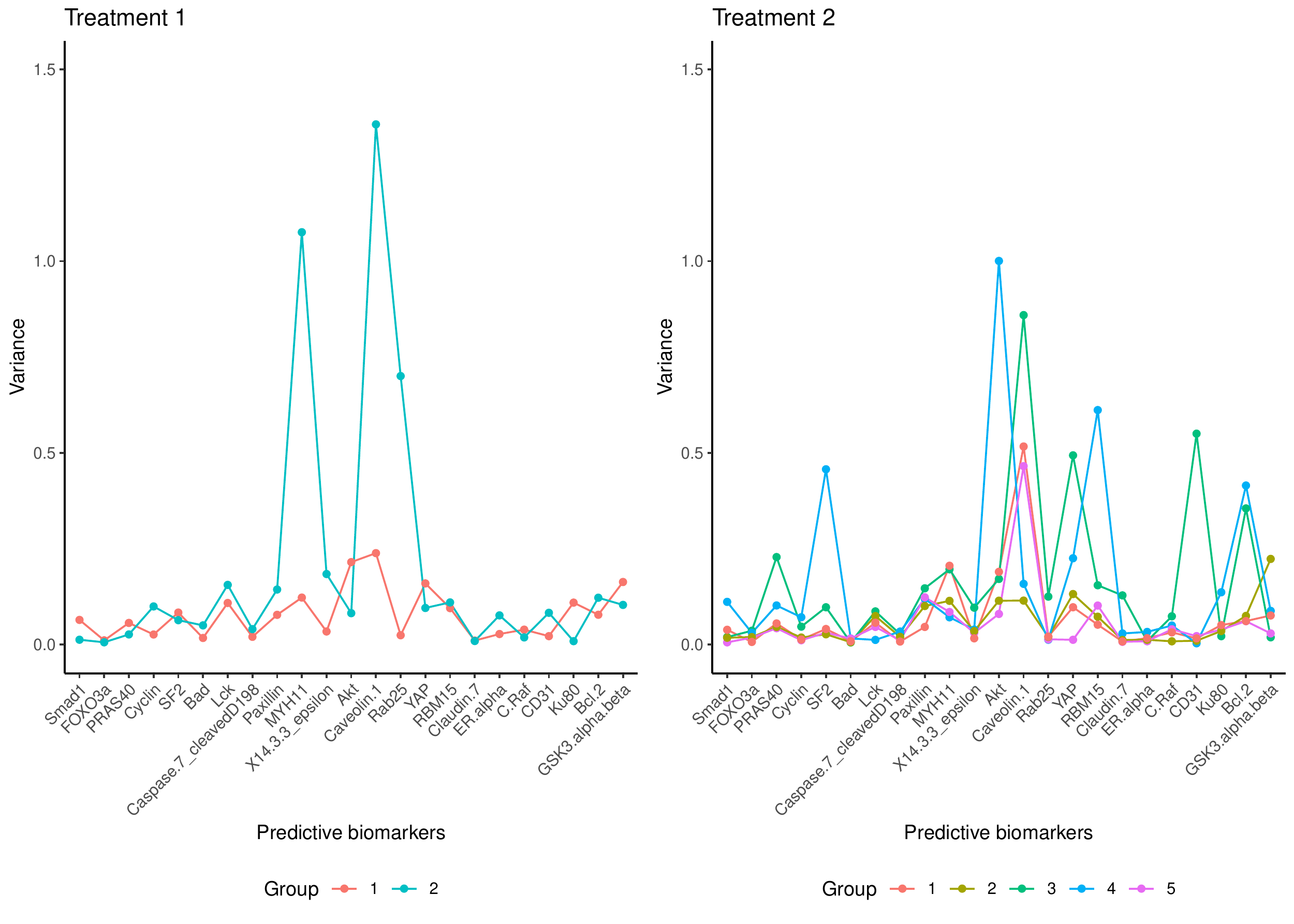}
    \caption{Group-specific variance of predictive biomarkers for patients that received Treatment 1 (left) and Treatment 2 (right). Groups T1G3, T1G5, and T2G9 are singleton and are not displayed. }
    \label{fig:var_group}
\end{figure}

Our PPMx model provides homogeneous clusters in terms of predictive covariates so that the resulting clustering represents a compact data partitioning. It substantially reduces the within-group variance for each predictive covariate, as displayed in Figure \ref{fig:var_group}. Predictive covariates are standardized, so it is clear that there is a dramatic reduction in within-group variance for most of the predictive biomarkers. Some proteins still show a pronounced variance. Since it happens mainly in T1G6, T2G5, and T2G10, it is likely due to the small number of patients clustered together in these groups (2, 2, and 3 patients, respectively). 

\section{Additional Simulations for LGG data}
In order to assess the robustness of the results for LGG data analysis, we run additional analysis. The results of the simulation study are reported in Table \ref{tab:omegarob}. Since the number of correctly predicted outcome (NPC) does not depend on the weights, we report here only the Empirical Summary Measure (ESM). 

    \begin{table}[htbp]
    \centering
    \begin{tabular}{c|*{5}{c}}
      \toprule
      $\omega_2$ & 20 & 40 & 60 & 80 \\
      \midrule
      pam-bp & 0.0701 & 0.0553 & 0.0401 & 0.0388 \\
      km-bp & 0.0295 & 0.0384 & 0.0335 & 0.0199 \\
      hc-bp & 0.0770 & 0.0285 & 0.0170 & 0.0185 \\
      dm-int & 0.0746 & 0.0746 & 0.0970 & \bf{0.0876} \\
      t-ppmx & \bf{0.1015} & \bf{0.1008} & \bf{0.1020} & 0.0751 \\
      \bottomrule
    \end{tabular}
    \caption{Results of 10-fold cross-validation to obtain the ESM for different values of the weight for the partial response, $\omega_2$. The weights for Progressive Disease and Complete Response are fixed at $\omega_1=0$ and $\omega_3=100$, respectively. The best performance for each $\omega_2$ value is highlighted in bold.}
    \label{tab:omegarob}
    \end{table}

    We find that the method is robust to different elicitation of $\omega_2$, and that the results are consistent with those reported in the paper.

    To evaluate the impact of combining PR and SD, we performed an analysis of LGG data considering the four RECIST categories for the response. The results of this analysis are presented in Table \ref{tab:catrob}. We evaluate the number of correctly predicted outcome (NPC) and the Empirical Summary Measure (ESM). Note that ESM is based on only two response categories, responders (CR) and non-responders (PD + PR + SD), whereas we used all four levels of the ordinal outcome in the data analysis and to implement personalized treatment selection.
    \begin{table}[htbp]
    \centering
    \begin{tabular}{c|cc}
      \toprule
       & $NPC$ & $ESM$ ($>3$)\\
      \midrule
      pam-bp & 51 & 0.0809 \\
      km-bp & 53 & 0.0134 \\
      hc-bp & 53 & 0.0582 \\
      dm-int & \bf{54} & 0.0840 \\
      t-ppmx & 50 & \bf{0.1259} \\
      \bottomrule
    \end{tabular}
  \caption{Predictive performance with four RECIST categories: metrics were obtained using 10-fold cross-validation. To compute ESM, the fourth RECIST category (Complete Response) was considered as the response. The best performance for each index is highlighted in bold.}
  \label{tab:catrob}
  \end{table}

  Compared to the results obtained using three categories for the outcome, we observed a general improvement for all methods in terms of ESM. While the two-step methods proposed by \cite{ma2019bayesian} showed a slight improvement in NPC, dm-int and t-ppmx exhibited a deterioration. However, this additional evaluation confirmed that our method outperforms all competitors in treatment selection. 

\section{Alternative Likelihoods}

We would like to acknowledge that beyond the Dirichlet-multinomial likelihood, there are alternative functions worth considering. For instance, the multinomial logit model presents a viable option that might yield comparable predictive performance. However, it is important to note that interpreting the multinomial logit parameters can be less straightforward since they represent log odds ratios with respect to a specific baseline response level.
Furthermore, given the ordinal nature of our response variable, ordinal regression emerges as a compelling and valid choice. This model allows us to effectively accommodate the ordered categories while maintaining the interpretability of results in terms of the ordinal relationship between response levels. Notably, in a study by \cite{page2021discovering}, an ordinal response for a PPMx model has been considered by assuming it to be a discretization of a continuous, real-valued latent score. To achieve this, fixed thresholds were introduced to create the ordinal categories. However, we recognize that this strategy has inherent limitations, as the outcomes may be sensitive to the choice of cutoff values. To address this concern, a potential solution could involve estimating the latent cutoffs as unknown parameters, which could lead to more robust and reliable results.

\bibliographystyle{apalike} 
\bibliography{biblio}

\begin{thebibliography}{}

\bibitem[Agresti, 2019]{agresti2019introduction}
Agresti, A. (2019).
\newblock {\em An introduction to categorical data analysis}.
\newblock John Wiley \& Sons.

\bibitem[Argiento et~al., 2016]{argiento2016blocked}
Argiento, R., Bianchini, I., and Guglielmi, A. (2016).
\newblock A blocked gibbs sampler for ngg-mixture models via a priori
  truncation.
\newblock {\em Statistics and Computing}, 26(3):641--661.

\bibitem[Argiento et~al., 2022]{argiento2022clustering}
Argiento, R., Corradin, R., Guglielmi, A., and Lanzarone, E. (2022).
\newblock Clustering blood donors via mixtures of product partition models with
  covariates.
\newblock {\em arXiv}.

\bibitem[Argiento and De~Iorio, 2022]{argiento2022infinity}
Argiento, R. and De~Iorio, M. (2022).
\newblock Is infinity that far? a bayesian nonparametric perspective of finite
  mixture models.
\newblock {\em The Annals of Statistics}, 50(5):2641--2663.

\bibitem[Barcella et~al., 2017]{barcella2017comparative}
Barcella, W., De~Iorio, M., and Baio, G. (2017).
\newblock A comparative review of variable selection techniques for covariate
  dependent dirichlet process mixture models.
\newblock {\em Canadian Journal of Statistics}, 45(3):254--273.

\bibitem[Bedard et~al., 2013]{bedard2013tumour}
Bedard, P.~L., Hansen, A.~R., Ratain, M.~J., and Siu, L.~L. (2013).
\newblock Tumour heterogeneity in the clinic.
\newblock {\em Nature}, 501(7467):355--364.

\bibitem[Bonetti and Gelber, 2000]{bonetti2000graphical}
Bonetti, M. and Gelber, R.~D. (2000).
\newblock A graphical method to assess treatment--covariate interactions using
  the cox model on subsets of the data.
\newblock {\em Statistics in Medicine}, 19(19):2595--2609.

\bibitem[Carvalho et~al., 2010]{carvalho2010horseshoe}
Carvalho, C.~M., Polson, N.~G., and Scott, J.~G. (2010).
\newblock The horseshoe estimator for sparse signals.
\newblock {\em Biometrika}, 97(2):465--480.

\bibitem[Chen and Li, 2013]{chen2013variable}
Chen, J. and Li, H. (2013).
\newblock Variable selection for sparse dirichlet-multinomial regression with
  an application to microbiome data analysis.
\newblock {\em The Annals of Applied Statistics}, 7(1):418--442.

\bibitem[Christensen et~al., 2011]{christensen2011bayesian}
Christensen, R., Johnson, W., Branscum, A., and Hanson, T.~E. (2011).
\newblock {\em Bayesian ideas and data analysis: an introduction for scientists
  and statisticians}.
\newblock CRC press.

\bibitem[Claus et~al., 2015]{claus2015survival}
Claus, E.~B., Walsh, K.~M., Wiencke, J.~K., Molinaro, A.~M., Wiemels, J.~L.,
  Schildkraut, J.~M., Bondy, M.~L., Berger, M., Jenkins, R., and Wrensch, M.
  (2015).
\newblock Survival and low-grade glioma: the emergence of genetic information.
\newblock {\em Neurosurgical Focus}, 38(1):E6.

\bibitem[Corsini and Viroli, 2022]{corsini2022dealing}
Corsini, N. and Viroli, C. (2022).
\newblock Dealing with overdispersion in multivariate count data.
\newblock {\em Computational Statistics \& Data Analysis}, 170:107447.

\bibitem[De~Blasi et~al., 2013]{de2013gibbs}
De~Blasi, P., Favaro, S., Lijoi, A., Mena, R.~H., Pr{\"u}nster, I., and
  Ruggiero, M. (2013).
\newblock Are gibbs-type priors the most natural generalization of the
  dirichlet process?
\newblock {\em IEEE Transactions on Pattern Analysis and Machine Intelligence},
  37(2):212--229.

\bibitem[Favaro et~al., 2013]{favaro2013mcmc}
Favaro, S., Teh, Y.~W., et~al. (2013).
\newblock Mcmc for normalized random measure mixture models.
\newblock {\em Statistical Science}, 28(3):335--359.

\bibitem[Gelman and Rubin, 1992]{gelman1992inference}
Gelman, A. and Rubin, D.~B. (1992).
\newblock Inference from iterative simulation using multiple sequences.
\newblock {\em Statistical Science}, pages 457--472.

\bibitem[Gnedin and Pitman, 2006]{gnedin2006exchangeable}
Gnedin, A. and Pitman, J. (2006).
\newblock Exchangeable gibbs partitions and stirling triangles.
\newblock {\em Journal of Mathematical Sciences}, 138(3):5674--5685.

\bibitem[Golub et~al., 1999]{golub1999molecular}
Golub, T.~R., Slonim, D.~K., Tamayo, P., Huard, C., Gaasenbeek, M., Mesirov,
  J.~P., Coller, H., Loh, M.~L., Downing, J.~R., Caligiuri, M.~A., et~al.
  (1999).
\newblock Molecular classification of cancer: class discovery and class
  prediction by gene expression monitoring.
\newblock {\em Science}, 286(5439):531--537.

\bibitem[Goodenberger and Jenkins, 2012]{goodenberger2012genetics}
Goodenberger, M.~L. and Jenkins, R.~B. (2012).
\newblock Genetics of adult glioma.
\newblock {\em Cancer Genetics}, 205(12):613--621.

\bibitem[Hartigan, 1990]{hartigan1990partition}
Hartigan, J.~A. (1990).
\newblock Partition models.
\newblock {\em Communications in Statistics-Theory and Methods},
  19(8):2745--2756.

\bibitem[Imai et~al., 2008]{imai2008misunderstandings}
Imai, K., King, G., and Stuart, E.~A. (2008).
\newblock Misunderstandings between experimentalists and observationalists
  about causal inference.
\newblock {\em Journal of the royal statistical society: series A (statistics
  in society)}, 171(2):481--502.

\bibitem[Ius et~al., 2018]{ius2018nf}
Ius, T., Ciani, Y., Ruaro, M.~E., Isola, M., Sorrentino, M., Bulfoni, M.,
  Candotti, V., Correcig, C., Bourkoula, E., Manini, I., et~al. (2018).
\newblock An nf-$\kappa$b signature predicts low-grade glioma prognosis: A
  precision medicine approach based on patient-derived stem cells.
\newblock {\em Neuro-oncology}, 20(6):776--787.

\bibitem[Kosorok and Laber, 2019]{kosorok2019precision}
Kosorok, M.~R. and Laber, E.~B. (2019).
\newblock Precision medicine.
\newblock {\em Annual Review of Statistics and its Application}, 6:263--286.

\bibitem[Lee et~al., 2022]{lee202utility}
Lee, J., Thall, P.~F., Lim, B., and Msaouel, P. (2022).
\newblock Utility-based bayesian personalized treatment selection for advanced
  breast cancer.
\newblock {\em Journal of the Royal Statistical Society: Series C (Applied
  Statistics)}.

\bibitem[Li et~al., 2020]{li2020downregulation}
Li, S.-Z., Hu, Y.-Y., Zhao, J.-L., Zang, J., Fei, Z., Han, H., and Qin, H.-Y.
  (2020).
\newblock Downregulation of fhl1 protein in glioma inhibits tumor growth
  through pi3k/akt signaling.
\newblock {\em Oncology Letters}, 19(6):3781--3788.

\bibitem[Lijoi et~al., 2007]{lijoi2007controlling}
Lijoi, A., Mena, R.~H., and Pr{\"u}nster, I. (2007).
\newblock Controlling the reinforcement in bayesian non-parametric mixture
  models.
\newblock {\em Journal of the Royal Statistical Society: Series B (Statistical
  Methodology)}, 69(4):715--740.

\bibitem[Ma et~al., 2015]{ma2015statistical}
Ma, J., Hobbs, B.~P., and Stingo, F.~C. (2015).
\newblock Statistical methods for establishing personalized treatment rules in
  oncology.
\newblock {\em BioMed Research International}, 2015.

\bibitem[Ma et~al., 2018]{ma2018integrating}
Ma, J., Hobbs, B.~P., and Stingo, F.~C. (2018).
\newblock Integrating genomic signatures for treatment selection with bayesian
  predictive failure time models.
\newblock {\em Statistical Methods in Medical Research}, 27(7):2093--2113.

\bibitem[Ma et~al., 2016]{ma2016bayesian}
Ma, J., Stingo, F.~C., and Hobbs, B.~P. (2016).
\newblock Bayesian predictive modeling for genomic based personalized treatment
  selection.
\newblock {\em Biometrics}, 72(2):575--583.

\bibitem[Ma et~al., 2019]{ma2019bayesian}
Ma, J., Stingo, F.~C., and Hobbs, B.~P. (2019).
\newblock Bayesian personalized treatment selection strategies that integrate
  predictive with prognostic determinants.
\newblock {\em Biometrical Journal}, 61(4):902--917.

\bibitem[Mills et~al., 2011]{mills2011emerging}
Mills, C.~N., Nowsheen, S., Bonner, J.~A., and Yang, E.~S. (2011).
\newblock Emerging roles of glycogen synthase kinase 3 in the treatment of
  brain tumors.
\newblock {\em Frontiers in Molecular Neuroscience}, 4:47.

\bibitem[Monti et~al., 2003]{monti2003consensus}
Monti, S., Tamayo, P., Mesirov, J., and Golub, T. (2003).
\newblock Consensus clustering: a resampling-based method for class discovery
  and visualization of gene expression microarray data.
\newblock {\em Machine learning}, 52(1):91--118.

\bibitem[M{\"u}ller et~al., 2011]{muller2011product}
M{\"u}ller, P., Quintana, F., and Rosner, G.~L. (2011).
\newblock A product partition model with regression on covariates.
\newblock {\em Journal of Computational and Graphical Statistics},
  20(1):260--278.

\bibitem[Neal, 2000]{neal2000markov}
Neal, R.~M. (2000).
\newblock Markov chain sampling methods for dirichlet process mixture models.
\newblock {\em Journal of Computational and Graphical Statistics},
  9(2):249--265.

\bibitem[Neal, 2003]{neal2003slice}
Neal, R.~M. (2003).
\newblock Slice sampling.
\newblock {\em The Annals of Statistics}, 31(3):705--767.

\bibitem[Olar and Sulman, 2015]{olar2015molecular}
Olar, A. and Sulman, E.~P. (2015).
\newblock Molecular markers in low-grade glioma—toward tumor
  reclassification.
\newblock In {\em Seminars in Radiation Oncology}, volume~25, pages 155--163.
  Elsevier.

\bibitem[Page and Quintana, 2015]{page2015predictions}
Page, G.~L. and Quintana, F.~A. (2015).
\newblock Predictions based on the clustering of heterogeneous functions via
  shape and subject-specific covariates.
\newblock {\em Bayesian Analysis}, 10(2):379--410.

\bibitem[Page and Quintana, 2016]{page2016spatial}
Page, G.~L. and Quintana, F.~A. (2016).
\newblock Spatial product partition models.
\newblock {\em Bayesian Analysis}, 11(1):265--298.

\bibitem[Page and Quintana, 2018]{page2018calibrating}
Page, G.~L. and Quintana, F.~A. (2018).
\newblock Calibrating covariate informed product partition models.
\newblock {\em Statistics and Computing}, 28(5):1009--1031.

\bibitem[Page et~al., 2021]{page2021discovering}
Page, G.~L., Quintana, F.~A., and Rosner, G.~L. (2021).
\newblock {Discovering interactions using covariate informed random partition
  models}.
\newblock {\em The Annals of Applied Statistics}, 15(1):1 -- 21.

\bibitem[Pocock et~al., 2002]{pocock2002subgroup}
Pocock, S.~J., Assmann, S.~E., Enos, L.~E., and Kasten, L.~E. (2002).
\newblock Subgroup analysis, covariate adjustment and baseline comparisons in
  clinical trial reporting: current practiceand problems.
\newblock {\em Statistics in Medicine}, 21(19):2917--2930.

\bibitem[Polson et~al., 2014]{polson2014bayesian}
Polson, N.~G., Scott, J.~G., and Windle, J. (2014).
\newblock The bayesian bridge.
\newblock {\em Journal of the Royal Statistical Society: Series B (Statistical
  Methodology)}, 76(4):713--733.

\bibitem[Poux-M{\'e}dard et~al., 2021]{poux2021powered}
Poux-M{\'e}dard, G., Velcin, J., and Loudcher, S. (2021).
\newblock Powered dirichlet process for controlling the importance of"
  rich-get-richer" prior assumptions in bayesian clustering.
\newblock {\em arXiv preprint arXiv:2104.12485}.

\bibitem[Quintana and Iglesias, 2003]{quintana2003bayesian}
Quintana, F.~A. and Iglesias, P.~L. (2003).
\newblock Bayesian clustering and product partition models.
\newblock {\em Journal of the Royal Statistical Society: Series B (Statistical
  Methodology)}, 65(2):557--574.

\bibitem[Ripley et~al., 2013]{ripley2013package}
Ripley, B., Venables, B., Bates, D.~M., Hornik, K., Gebhardt, A., Firth, D.,
  and Ripley, M.~B. (2013).
\newblock Package ‘mass’.
\newblock {\em Cran r}, 538:113--120.

\bibitem[Simon, 2010]{simon2010clinical}
Simon, R. (2010).
\newblock Clinical trial designs for evaluating the medical utility of
  prognostic and predictive biomarkers in oncology.
\newblock {\em Personalized Medicine}, 7(1):33--47.

\bibitem[Song and Pepe, 2004]{song2004evaluating}
Song, X. and Pepe, M.~S. (2004).
\newblock Evaluating markers for selecting a patient's treatment.
\newblock {\em Biometrics}, 60(4):874--883.

\bibitem[Stuart et~al., 2011]{stuart2011matchit}
Stuart, E.~A., King, G., Imai, K., and Ho, D. (2011).
\newblock Matchit: nonparametric preprocessing for parametric causal inference.
\newblock {\em Journal of statistical software}.

\bibitem[Wade and Ghahramani, 2018]{wade2018bayesian}
Wade, S. and Ghahramani, Z. (2018).
\newblock Bayesian cluster analysis: Point estimation and credible balls (with
  discussion).
\newblock {\em Bayesian Analysis}, 13(2):559--626.

\bibitem[Zhang et~al., 2012]{zhang2012robust}
Zhang, B., Tsiatis, A.~A., Laber, E.~B., and Davidian, M. (2012).
\newblock A robust method for estimating optimal treatment regimes.
\newblock {\em Biometrics}, 68(4):1010--1018.

\bibitem[Zhao et~al., 2012]{zhao2012estimating}
Zhao, Y., Zeng, D., Rush, A.~J., and Kosorok, M.~R. (2012).
\newblock Estimating individualized treatment rules using outcome weighted
  learning.
\newblock {\em Journal of the American Statistical Association},
  107(499):1106--1118.

\end{thebibliography}
\end{document}